\documentclass[reprint,tikz,twocolumn,prl,superscriptaddress]{revtex4-1}
\AtBeginDocument{
\heavyrulewidth=.08em
\lightrulewidth=.05em
\cmidrulewidth=.03em
\belowrulesep=.65ex
\belowbottomsep=0pt
\aboverulesep=.4ex
\abovetopsep=0pt
\cmidrulesep=\doublerulesep
\cmidrulekern=.5em
\defaultaddspace=.5em
}

\usepackage{amsmath,amssymb}
\usepackage{graphicx}

\usepackage{float}
\usepackage[usenames,dvipsnames]{xcolor}
\usepackage{amsthm,comment}
\usepackage{booktabs}

\usepackage[printwatermark]{xwatermark}
\usepackage[export]{adjustbox}
\usepackage[section]{placeins}
\usepackage[caption=false]{subfig}
\usepackage[percent]{overpic}
\bibpunct{[}{]}{,}{n}{}{}
\definecolor{red1}{HTML}{FF4136}
\definecolor{green1}{HTML}{00802b}
\newcommand{\pdag}{{\phantom\dagger}}

\usepackage[hidelinks]{hyperref} 

\hypersetup{colorlinks=true,citecolor=Red,linkcolor=Blue,urlcolor=Blue}
\usepackage[export]{adjustbox}

\newcommand{\<}{\langle}
\renewcommand{\>}{\rangle}
\newcommand{\up}{\uparrow}
\newcommand{\dn}{\downarrow}
\renewcommand{\c}[1]{c_{#1}} 
\newcommand{\cd}[1]{c_{#1}^{\dagger}} 
\newcommand{\w}{\omega}
\newcommand{\Akw}[1]{A^{#1}{(k,\omega)}}
\newcommand{\Nkw}[1]{N^{#1}{(k,\omega)}}
\newcommand{\Lkw}[1]{L^{#1}{(k,\omega)}}
\newcommand{\Skw}[1]{S^{#1}{(k,\omega)}}

\begin{document}

\title{Fingerprints of an Exotic Orbital-Selective Mott Phase \\ 
in the Block Magnetic State of BaFe$_2$Se$_3$ Ladders}

\author{N. D. Patel}
\affiliation{Department of Physics and Astronomy, The University of Tennessee, Knoxville, Tennessee 37996, USA}
\affiliation{Materials Science and Technology Division, Oak Ridge National Laboratory, Oak Ridge, Tennessee 37831, USA}

\author{A. Nocera}
\affiliation{Department of Physics and Astronomy, The University of Tennessee, Knoxville, Tennessee 37996, USA}
\affiliation{Materials Science and Technology Division, Oak Ridge National Laboratory, Oak Ridge, Tennessee 37831, USA}

\author{G. Alvarez}
\affiliation{Computer Science \& Mathematics %
Division and Center for Nanophase Materials Sciences, Oak Ridge National Laboratory, %
 \mbox{Oak Ridge, Tennessee 37831}, USA}

\author{A. Moreo}
\affiliation{Department of Physics and Astronomy, The University of Tennessee, Knoxville, Tennessee 37996, USA}
\affiliation{Materials Science and Technology Division, Oak Ridge National Laboratory, Oak Ridge, Tennessee 37831, USA}

\author{S. Johnston}
\affiliation{Department of Physics and Astronomy, The University of Tennessee, Knoxville, Tennessee 37996, USA}
\affiliation{Joint Institute for Advanced Materials at The University of Tennessee, Knoxville, Tennessee 37996, USA}

\author{E. Dagotto}
\affiliation{Department of Physics and Astronomy, The University of Tennessee, Knoxville, Tennessee 37996, USA}
\affiliation{Materials Science and Technology Division, Oak Ridge National Laboratory, Oak Ridge, Tennessee 37831, USA}

\begin{abstract}
Resonant Inelastic X-Ray Scattering (RIXS) experiments on the iron-based ladder
BaFe$_2$Se$_3$ unveiled an unexpected two-peak structure associated with local
orbital ($dd$) excitations in a block-type antiferromagnetic phase. A mixed
character between correlated band-like and localized excitations was also
reported. Here, we use the density matrix renormalization group method to
calculate the momentum-resolved charge- and orbital-dynamical response
functions of a multi-orbital Hubbard chain. Remarkably, our results
qualitatively resemble the BaFe$_2$Se$_3$ RIXS data, while also capturing the
presence of long-range magnetic order as found in neutron scattering, {\it
only} when the model is in an exotic orbital-selective Mott phase (OSMP). In
the calculations, the experimentally observed zero-momentum transfer RIXS peaks
correspond to excitations between itinerant and Mott insulating orbitals. We
provide experimentally testable predictions for the momentum-resolved charge
and orbital dynamical structures, which can provide further insight into the
OSMP regime of BaFe$_2$Se$_3$. 
\end{abstract}

\maketitle

\begin{figure}[t]
\begin{center}
 \begin{overpic}[trim = 0cm 0.0cm 0mm 0mm,angle=0]{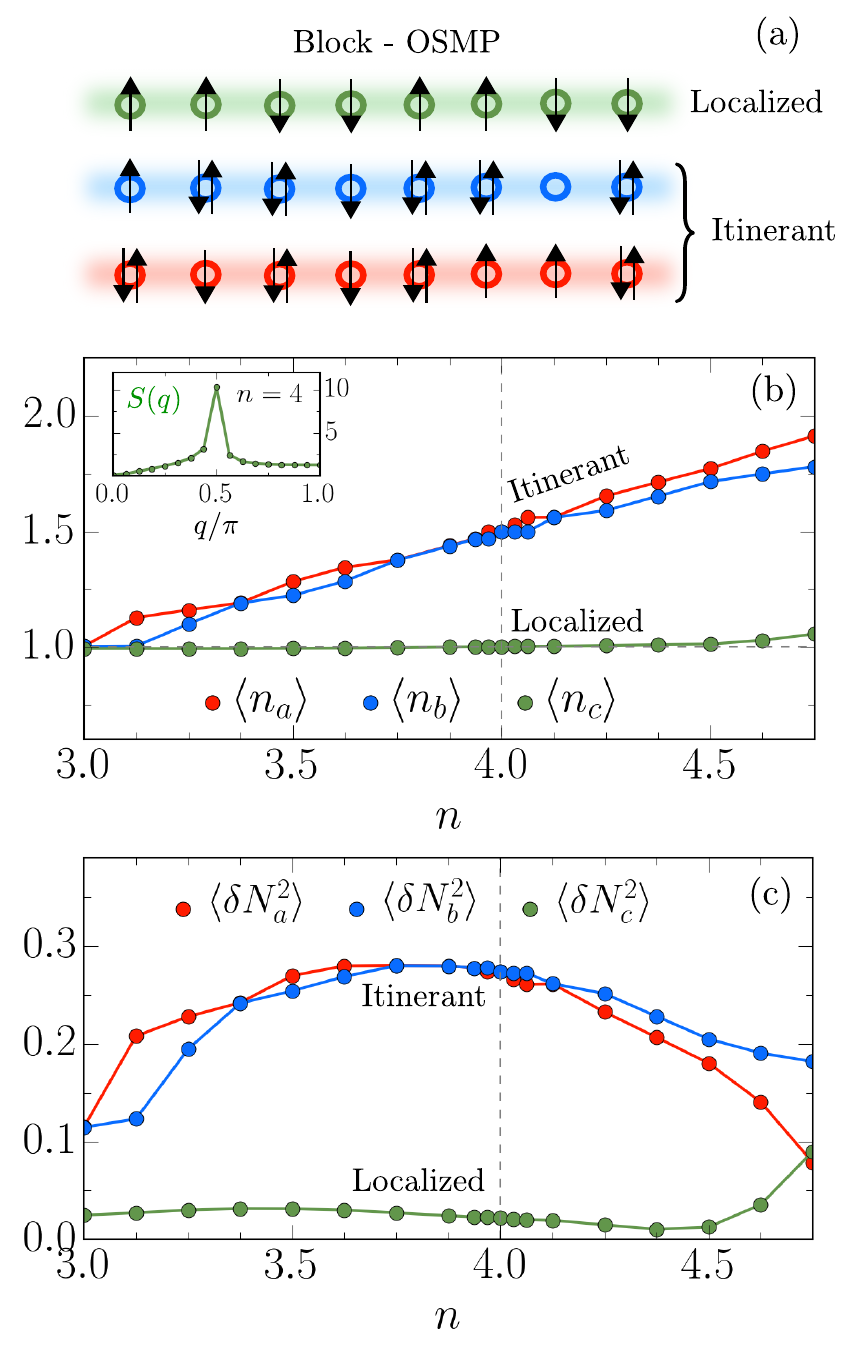}
\end{overpic}
\end{center}
\subfloat{\label{fig:1a}}
\subfloat{\label{fig:1b}}
\subfloat{\label{fig:1c}}
\vspace{-1.2cm}
\caption{(a) Sketch of the block OSMP state of focus in the present effort.
The upper orbital (green) is localized while the other two (blue, red) are itinerant. Illustration adapted from \cite{JacekSpinOSMP}.
(b) Average local occupation of each orbital vs overall electron density $n$, where $n=4$ represents the filling of $4$ electrons per site (using DMRG, 32 sites), which is a realistic situation when a three-orbital model is used. 
The inset shows the total static spin-structure factor with a peak at $q = \pi/2$, representing the block-type AFM order at $n=4$, with two spins ``up'', followed by two ``down'', and then a repeated periodicity, as illustrated in (a) upper orbital. (c) Average local charge fluctuations on each orbital vs $n$.
}
\label{fig:1}
\end{figure}

{\bf Introduction} --- The recent discovery of superconductivity in the two-leg
iron-based ladder compounds BaFe$_2$S$_3$ \cite{NatMatSC,PRLSC} and
BaFe$_2$Se$_3$ \cite{123SeSC1,*123SeSC2,*Pottgen1} opened an exciting new
branch of research: these materials are the first members of the iron
superconductors family
\cite{ironSCfamily1,*ironSCfamily2,nphys2438,DagottoRMP13} without iron layers,
the parent compounds are insulators \cite{NatMatSC,PRLSC}, and their
low-dimensionality allows for accurate theoretical treatment \cite{WhiteDMRG1,
DMRGreview, white2005, WhiteWFT, GonzaloDMRGpp}.  For this reason, iron-ladder
compounds were the focus of many recent experimental and theoretical studies
\cite{NatMatSC,PRLSC,123SeSC1,PatelBaFe2S3,
NambuBaFeSe,Patel2OrbChain,CaronBaFeSe, FeLadder2,FeLadder5,FeLadder7,
NeutronOSMP, MonneyRIXS, PEOhgushi}.  In particular, inelastic neutron
scattering data  \cite{NeutronOSMP} is compatible with the exciting idea that
BaFe$_2$Se$_3$ is in an orbital-selective Mott phase (OSMP) at ambient pressure
\cite{georges,yu-si, JulianOSMP}, where one orbital is localized with a Mott
gap while the others are gapless and itinerant (see Fig.~\ref{fig:1a}).
Moreover, this state displays an exotic magnetic order involving $2\times 2$
ferromagnetic blocks that are antiferromagnetically staggered. 

The intuitive origin of the block magnetic state admits multiple descriptions.
Within Hartree-Fock treatments~\cite{FeLadder7}, the ``block'' structure arises
from magnetic frustration because the system is located in parameter space
between a ferromagnetic state, induced by double-exchange at large Hund
coupling, and a staggered antiferromagnetic state, induced by superexchange at
small Hund coupling. Alternatively, recent efforts based on an itinerant
perspective predicted that the metallic orbitals are the ``drivers'' while the
localized spins are the ``passengers'', with the block structure arising from
Fermi surface nesting \cite{Jacekpreparation}. From the experimental
perspective, Resonant Inelastic X-ray Scattering (RIXS) and X-ray Photoemission
(XPS) studies report the presence of both localized and itinerant carriers in
BaFe$_2$Se$_3$,  also suggesting OSMP physics \cite{MonneyRIXS, PEOhgushi};
however, up to now, no sufficient theoretical study has been carried out to
support these indirect experimental claims of an OSMP ground state in
BaFe$_2$Se$_3$. Our primary goal is to fill this gap and provide evidence from
a theoretical perspective that indeed the ``123'' ladder is in an OSMP state. 

\begin{figure*}[t]
\begin{center}
 \begin{overpic}[trim = 0cm 0mm 0cm 0mm,width=0.99\textwidth,angle=0]{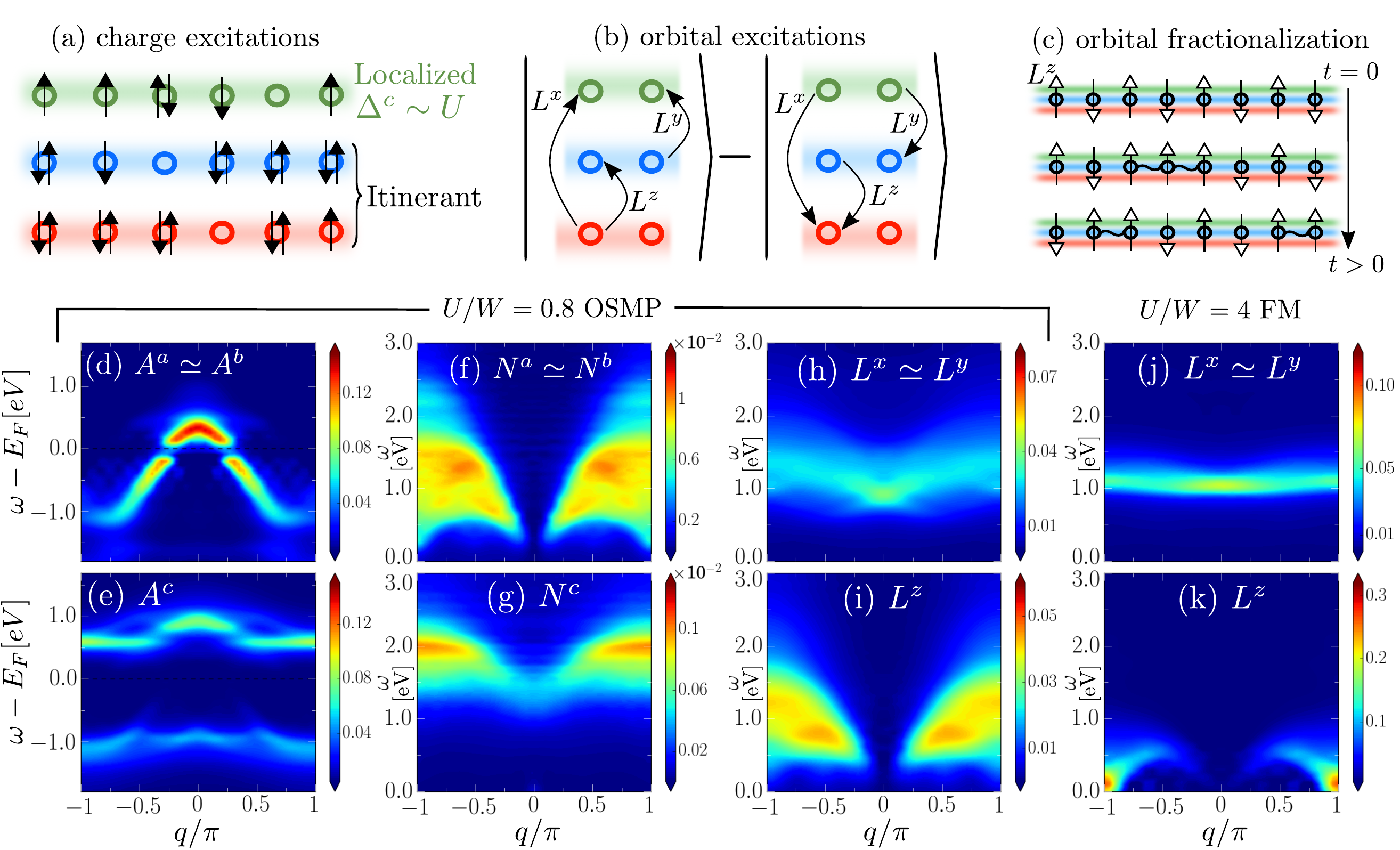}
\end{overpic}
\subfloat{\label{fig:2a}}
\subfloat{\label{fig:2b}}
\subfloat{\label{fig:2c}}
\subfloat{\label{fig:2d}}
\subfloat{\label{fig:2e}}
\subfloat{\label{fig:2f}}
\subfloat{\label{fig:2g}}
\subfloat{\label{fig:2h}}
\subfloat{\label{fig:2i}}
\subfloat{\label{fig:2j}}
\subfloat{\label{fig:2k}}
\end{center}
\vspace{-0.7cm}
\caption{Sketches of (a) charge and (b) orbital excitations  in a block OSMP state. (c) is a sketh of ``orbital fractionalization'' 
discussed in the FM insulator context below. Actual DMRG data shown correspond to the orbital-resolved (d-e) single-particle spectra, 
(f-g) charge excitations, and (h-i) orbital excitations spectra within the block magnetic OSMP state. 
$\Akw{a/b}$ has reduced but nonzero quasi-particle weight at the Fermi level ($E_F$) while $\Akw{c}$ 
has a characteristic Mott gap $\Delta_c \sim 1.5$~eV.
The charge excitations $\Nkw{a/b}$ have a gapless continuum at finite $\w$ 
while $\Nkw{c}$ has an excitation gap of $\Delta_c^{N} \sim 1.5$~eV. 
$\Lkw{x/y}$ represents inter-orbital excitations between orbitals $a/b$ and $c$
($d_{xz/yz} \leftrightarrow d_{xy}$) that are gapped with an excitation gap $\Delta^{L} \sim 0.7$-$0.9$~eV.
$\Lkw{z}$ are inter-orbital excitations between orbitals $a$ and $b$ 
($d_{xz} \leftrightarrow d_{yz}$). (j-k) Orbital excitations spectra within the 
FM insulator competing state. $\Lkw{x/y}$ has a gap $\sim 0.9$~eV and $\Lkw{z}$ is gapless with 
a robust $q=\pi$ peak indicating quasi-long range staggered orbital ordering.
Spectral functions are calculated 
using $\Delta \w = 0.05$~eV, broadening $\eta = 0.1$~eV, up to $1200$ DMRG states, 24 sites (d-g), 16 sites (h-k), 
and $8$ DMRG sweeps. 
}
\label{fig:2}
\end{figure*}

One of the first steps towards unveiling the characteristic  
excitations of an OSMP state is to calculate its single-particle spectral function $A(\bf{q},\w)$ 
and its intra- and inter-orbital dynamical spin $S(\bf{q},\w)$, 
charge $N(\bf{q},\w)$, and orbital $L(\bf{q},\w)$ structure factors. 
Recently, theoretical predictions for $S(\bf{q},\w)$ were presented \cite{JacekSpinOSMP} in a block antiferromagnetic (AFM) OSMP. 
Using determinantal quantum Monte Carlo and the maximum entropy method, the low temperature $A(\bf{q},\w)$ was also reported employing an approximation to a multi-orbital Hubbard model \cite{SLiOSMP}. 
{\it However,  $N(\bf{q},\w)$ and $L(\bf{q},\w)$, crucial for RIXS experiments, have not been addressed yet.} Therefore, here for the first time, we use the density matrix renormalization group (DMRG) technique to compute the momentum-resolved charge and orbital response functions (see Figs.~\ref{fig:2a} and Figs.~\ref{fig:2b}), as well as the single-particle spectral function (at zero temperature), of the block magnetic OSMP of a multi-orbital Hubbard chain. 

There are multiple reasons for focusing on a chain geometry as opposed to a ladder. First, a three-orbital Hubbard chain is already equivalent to a three-leg one-orbital ladder; thus a three-orbital Hubbard ladder maps onto a six-leg one-orbital ladder, which is challenging even with DMRG. Second, in the usual ``snake'' geometry of DMRG, interorbital hoppings mutate into long-distance hoppings in the one-orbital chain analog, here involving sites effectively separated by eight lattice spacings. Such long-range hoppings compromise the accuracy of DMRG. Finally, RIXS in the eV scale usually addresses local excitations, and there should not be much difference between chains and ladders, as recent efforts in neutron scattering using both chains and ladders showed \cite{JacekSpinOSMP}.

We compare our results against previously gathered RIXS
data for BaFe$_2$Se$_3$ \cite{MonneyRIXS}, which
measures magnetic, charge, and orbital excitations
simultaneously~\cite{RMPRIXS1, RIXS1, RIXS2, RIXS3, JohnstonNatureComm}. By calculating the orbital response functions of the competing paramagnetic metal (PM) and ferromagnetic (FM) insulator states, we show that block OSMP has a characteristic two-peak structure that is distinctive and in striking agreement with RIXS results on BaFe$_2$Se$_3$ \cite{MonneyRIXS}. Moreover, we identify the observed $dd$ peaks as orbital excitations between localized and
itinerant orbitals. Our study strongly suggests that the ground state of BaFe$_2$Se$_3$ is indeed an OSMP with block magnetic order. 

%
%

{\bf Results} --- Figure~1 plots the local charge occupation (panel b) and fluctuations (panel c) of each orbital vs the total electronic
density, which demonstrates that the OSMP
is stable and robust against variations in the hole and electron doping~\cite{rincon2}. 
First, we focus on the results for 
$4$ electrons per site ($n=4$), shown by the vertical dashed line 
in Fig.~\ref{fig:1}~(b,c) (note that $n=4$ in a three-orbital model is the analog of the realistic $n=6$ in a five-orbital model). At this density, orbital $c$ has one electron per site, and thus becomes Mott localized,
while the two other orbitals remain fractionally occupied, and thus metallic. The reason for this exotic behavior relies on the crystal-field splitting and different bandwidths of the orbitals: as the
Hubbard $U$ interaction opens a gap and shifts energy states ``up and down'' relative to the gap, it becomes energetically favorable for orbital $c$ 
to have a half-filled lower Hubbard band, which is formed by moving electrons from the other bands into $c$.  
The corresponding intra-orbital charge fluctuations $\< \delta N_\gamma^2 \> = \langle n_\gamma^2\rangle - \langle n_\gamma\rangle^2$ in orbital $c$ are significantly suppressed, compatible with a localized state, while charge fluctuations in  orbitals $a$ and $b$ remain finite, compatible with a metallic state. 
The inset of Fig.~\ref{fig:1b} establishes the magnetic order at this filling, and plots the spin structure factor at $n=4$, displaying the characteristic peak at $q = \pi/2$ that is in agreement with the experimentally observed non-trivial block-type AFM order. 
Note that orbital $c$ (representing the $d_{xy}$ orbital, see methods) has the standard 
characteristics of a Mott phase at $n=4$, where charge degrees of 
freedom are ``frozen'' (localized) and accompanied by well-formed local 
magnetic moments, $\mathbf{m}_c \sim 0.98\mu_B$.  In  contrast, the large local charge fluctuations of the other orbitals suggest a metallic behavior typical of itinerant electrons. 
The total on-site local moment at $n=4$ is $\mathbf{m}_{tot} \simeq 1.97\mu_B$, a robust value slightly larger than in experiments \cite{NeutronOSMP,CaronBaFeSe,NambuBaFeSe}.


%
%

 The coexistence of localized and itinerant carriers is also evident when examining the electronic density of each orbital vs the global filling near the commensurate value $n=4$ (Fig.~\ref{fig:1b}). 
 The linear behavior of $n_a$ and $n_b$ suggests a band-like picture, 
 while the robust plateau in $n_c$ extending over a wide range of doping ($n = [3:4.5]$) indicates a Mott gap in the single-particle spectral function of orbital $c$. Since all these results provide ample evidence that we correctly capture the block-magnetic properties of BaFe$_2$Se$_3$, henceforth we fix the filling $n=4$ to address the features 
of the block OSMP in the RIXS dynamical excitations, the focus of
our publication.  

\begin{figure}[t]
\begin{center}
 \begin{overpic}[trim = 0cm 0cm 0.0cm 0cm,angle=0]{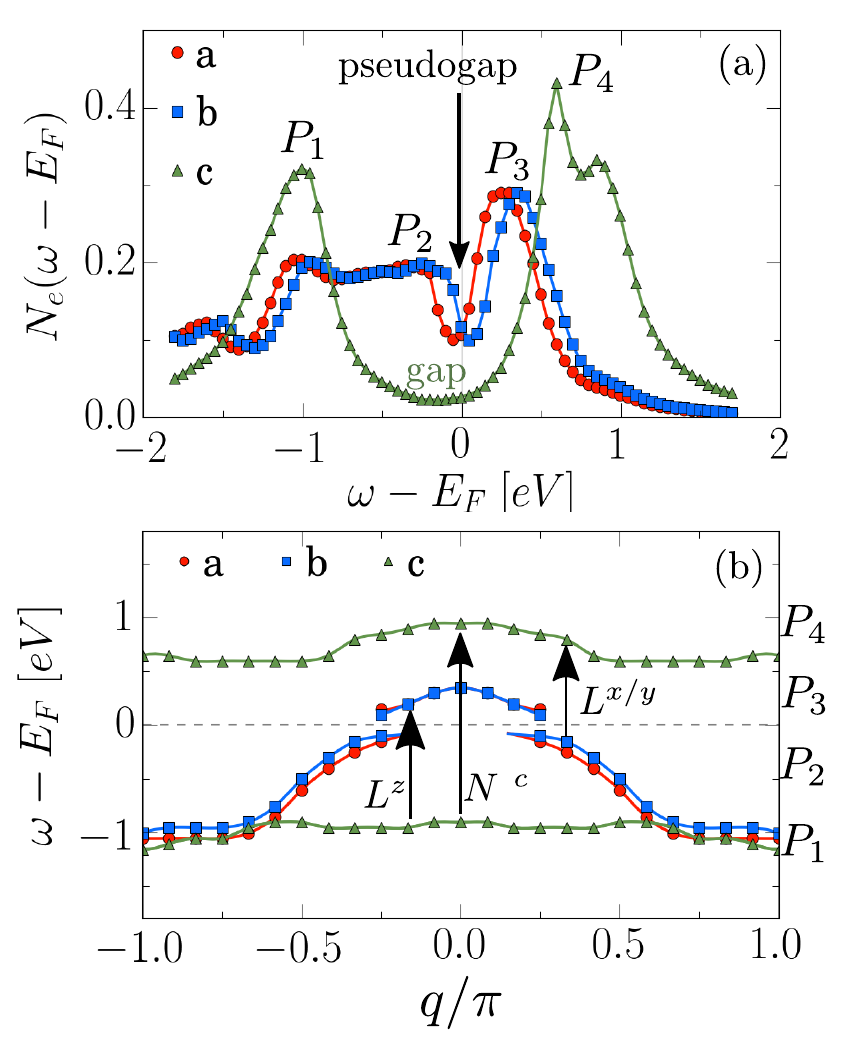}
\end{overpic}
\end{center}
\subfloat{\label{fig:3a}}
\subfloat{\label{fig:3b}}
\vspace{-1.2cm}
\caption{Orbital-resolved (a) density of states, and 
 (b) poles of the total single-particle spectra (Figs.~\ref{fig:2d} and \ref{fig:2e})
 of the block-antiferromagnetic  OSMP (using DMRG with 24 sites). 
}
\label{fig:3}
\end{figure}

{\bf Dynamical spectra of the block OSMP} --- Figures~\ref{fig:2d} and \ref{fig:2e} show the orbital-resolved single-particle spectral functions. 
Orbitals $a$ and $b$ have a finite quasi-particle weight at the Fermi level $E_F$, as also shown in the density-of-states in Figure~\ref{fig:3a}, confirming once again the itinerant nature of electrons in these orbitals. 
Moreover, we find that the orbital $c$ has a gap of value
$\Delta_{c} \sim 1.5~\text{eV}$ \cite{footnote-cgap} in $\Akw{c}$ (Figs.~\ref{fig:2e} and \ref{fig:3a}), 
compatible with our analysis in Figs.~\ref{fig:1b} and \ref{fig:1c}. 
The spectral function demonstrates again the coexistence of gapped Mott (localized) and gapless itinerant carriers and agrees well with an earlier study of a similar model \cite{SLiOSMP}. 
Note also the presence of a pseudogap in the itinerant $d_{xz/yz}$ orbitals (not captured by earlier work~\cite{SLiOSMP}). 
It is likely that this pseudogap originates from correlations between the block ordered spins of Mott orbital $c$  
and itinerant electrons in orbitals $a$ and $b$ (Fig.~\ref{fig:3a}), since they are not decoupled from one another. 
We also highlight that the Fermi momentum of orbitals $a$ and $b$ 
is approximately $q_F \simeq \pi/4$, leading to scattering at $2q_F \simeq \pi/2$ that is comparable 
to the spin-structure factor peak at $q = \pi/2$ corresponding to the 
block AFM order \cite{Jacekpreparation}. It is conceivable that the block phase of localized spins of the 
Mott orbital $c$ is driven by the Fermi nesting of itinerant orbitals, as
in manganites and heavy-fermion systems \cite{Jacekpreparation}.
In fact, recent experimental \cite{NeutronOSMP} and theoretical \cite{JacekSpinOSMP} work have
confirmed that the magnetic excitations in BaFe$_2$Se$_3$ cannot be 
fully described using an effective Heisenberg model, 
highlighting the role of the other degrees of freedom in this material that we are focusing on here. 


To arrive to our main conclusions, 
we now calculate the charge (Figs. \ref{fig:2f} and \ref{fig:2g}) 
and orbital (Figs.~\ref{fig:2h} and \ref{fig:2i}) dynamics of the block-antiferromagnetic OSMP.  
Since the low-energy spin-dynamics have already been reported in the literature, we only focus on the high-energy charge and orbital dynamics that are more relevant to RIXS measurements at energy losses above 1 eV. 
As with $\Akw{a/b}$, the charge excitations of the itinerant orbitals display a 
{\it gapless} continuum (Fig.~\ref{fig:2f}) because charge fluctuations can propagate freely in a metallic system (at fixed-energy transfer $\w$ many states allow for charge fluctuations without an energy cost). In contrast to  this, the charge excitations of the Mott orbital $c$ display a
gap $\Delta^{N}_{c} \sim 1.5~\text{eV}$ (Fig.~\ref{fig:2g}), and $\Nkw{c}$ shows incoherent charge excitations, compatible 
with doubly occupied orbitals $d_{xy}$ (i.e. $c$) configurations 
with an energy proportional to the Mott gap (Fig.~\ref{fig:2a}).
We can contrast our results with the much studied one-orbital two-leg ladder Hubbard model, where charge excitations are gapped in the half-filled Mott phase but display a gapless continuum in the metallic phase away from half-filling \cite{AlbertoRPAvsFLEX}. Our results show the {\it simultaneous} presence of localized and itinerant fermions in our multi-orbital model.


Consider now the $q$-resolved orbital excitations  in the block OSMP.
Figure~\ref{fig:2h} ($\Lkw{x/y}$) shows the excitations 
from (to) the itinerant orbitals $a/b$ to (from) the localized orbital $c$. 
Figure~\ref{fig:2i} displays a gapless continuum in orbital excitations, $\Lkw{z}$, between the itinerant orbitals $a$ and $b$ 
(see illustration Fig.~\ref{fig:2b}). 
 These dynamical spectra can be  
 understood using the density-of-states (Fig.~\ref{fig:3a}), and 
 poles of the single-particle spectral function of the OSMP (Fig.~\ref{fig:3b}). 
Figure~\ref{fig:3a} shows that electron scattering from 
$P_2 \rightarrow P_4$ requires 
an energy transfer $\sim 0.9$~eV (vertical
arrow in Fig.~\ref{fig:3b}). 
This scattering creates a gapped response peak in the orbital excitations at $q \simeq 0$ and 
$\w \simeq 0.9~\text{eV}$ (Fig.~\ref{fig:2h}). 
The charge gap of
orbital $c$ in $\Nkw{c}$ can also be understood via the scattering between points $P_1$ and $P_4$ of 
the density-of-states, shown by another vertical arrow in Fig.~\ref{fig:3b}. 
Additionally, the gapless continuum in $\Lkw{z}$ is very similar to the itinerant charge excitations, 
$\Nkw{a/b}$ because only the (almost degenerate) itinerant orbitals contribute in the calculations of $\Lkw{z}$. 

%
%

{\bf Dynamical spectra of the FM insulator} --- To identify features of orbital excitations in the block OSMP that are unique, we also show the $q$-resolved orbital excitations in the competing 
FM insulator known to exist  at $U/W=4$,\cite{JulianOSMP} see
Figs.~\ref{fig:2j} and~\ref{fig:2k}. 
In this FM phase, which is not yet known to be stable experimentally, 
theoretical calculations show that the single-particle spectral functions of all the orbitals have a Mott gap~\cite{SLiOSMP}. Figures~\ref{fig:2j} and~\ref{fig:2k} show that 
the corresponding $\Lkw{x/y}$ and $\Lkw{z}$ display clearly 
sharper excitations in the FM phase in comparison to the block OSMP. 
Remarkably, $\Lkw{z}$ is gapless with a $\w \simeq 0$ peak at $q = \pi$, implying quasi-long-range antiferro-{\it orbital} order in the 
ground state, as discussed before~\cite{SLiOSMP}. 
In fact, the $\Lkw{z}$ of this FM phase resembles 
the $\Skw{}$ of the spin-$1/2$ Heisenberg chain with a two-spinon continuum. 
{\it By analogy, we conjecture that the 
$\Lkw{z}$ of the FM phase may denote the existence of fractionalized orbital excitations} \cite{schlappa2012spin}
(see illustration Fig.~\ref{fig:2c}). 
Orbital and spin excitations were also studied recently in the context of Kugel-Khomskii models \cite{orbext1,orbext2,orbext3}. 
Here, we show novel results of orbital fractionalization in a FM 
insulator using a general multi-orbital fermionic Hamiltonian.

\begin{figure}[ht]
\begin{center}
\begin{overpic}[trim = 0cm 0cm 0cm 0cm,height=0.84\textwidth,angle=0]{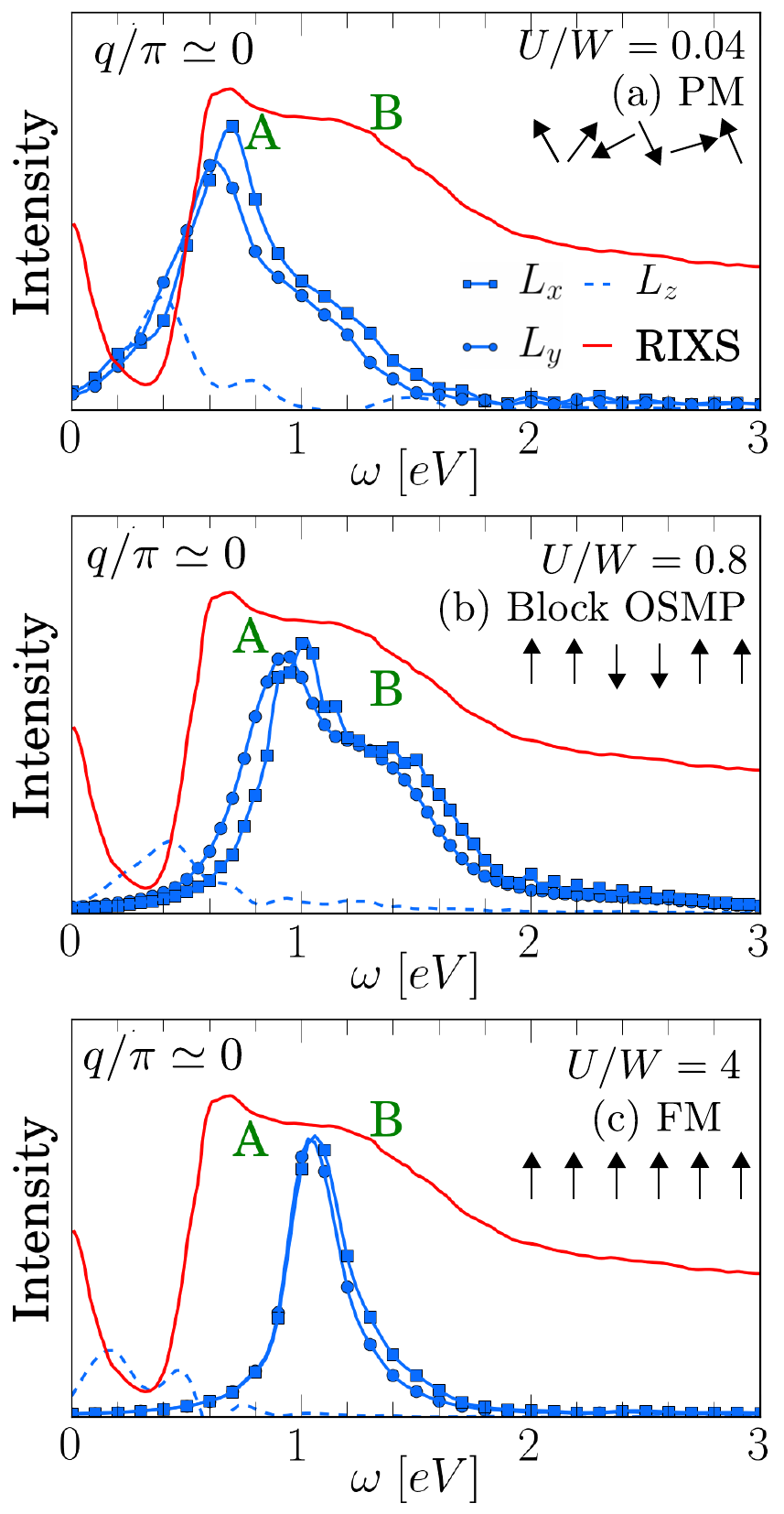}
\end{overpic}\vspace{-0.11cm} 
\end{center}
\subfloat{\label{fig:4a}}
\subfloat{\label{fig:4b}}
\subfloat{\label{fig:4c}}
\vspace{-0.8cm}
\caption{Orbital excitations spectra (blue) at zero-momentum transfer
corresponding to the three competing states, namely (a) $U/W=0.04$ (PM phase), (b) $U/W=0.8$ (block OSMP),  
and (c) $U/W=4$ (FM insulator phase), using DMRG and 16 sites. For comparison, 
each panel shows the experimental RIXS data (red) at the Fe-$L_3$ edge~\cite{MonneyRIXS}, 
where peaks $A$ and $B$ are $dd$ excitations. Peak $A$ is caused by the $P_2 \rightarrow P_4$ excitations of Fig.~\ref{fig:3b}, while
the subdominant peak $B$ are believed to be caused by the $P_1 \rightarrow P_3$ excitations of the same figure.
}
\label{fig:4}
\end{figure}

{\bf Discussion} ---  Our main result is in Fig.~\ref{fig:4}, where we
compare the calculated orbital dynamical 
response vs RIXS Fe-$L_3$ edge experimental data for BaFe$_2$Se$_3$ at zero-momentum transfer \cite{MonneyRIXS}. 
Relating the full RIXS intensity to dynamical structure
factors is not straightforward \cite{RIXS3, JiaPRX}. For example, for the single-band Hubbard model, the RIXS intensity can be directly related to dynamical structure factors only in limiting cases. To further complicate matters, 
the experimental effort \cite{MonneyRIXS} did not report the exact momentum transfer of
the experiment. While the scattering geometry employed nominally probes $q = 0$ excitations 
along the chain, there were difficulties in aligning the chain orientation during beam time \cite{Monneyprivate} 
and some finite $q$ excitations may have been mixed into the spectra.   
Additionally, RIXS measures excitations of many different channels (e.g. spin,
charge, and orbital), making it difficult to differentiate between those
distinct channels. However, comparisons with the dynamical response functions
can still provide at least {\it qualitative} insights into
the dominant features and energy scales of RIXS experiments. Moreover, a merit
of our effort is that we calculate excitations for each channel {\it
separately}, and can, therefore, provide detailed predictions for future
experiments. Until additional data become available
our focus on RIXS experiments at $q = 0$ momentum transfer~\cite{MonneyRIXS} is
the natural avenue to pursue.

%
%

%
%

At zero energy transfer, there is an ``elastic'' peak commonly observed in RIXS experiments. We note that magnon contributions to the RIXS data generally occur at energy losses $\w<0.2$ in the Fe-based superconductors, and, 
therefore, are hidden in the elastic contribution in the RIXS experiment \cite{NeutronOSMP,JacekSpinOSMP,MonneyRIXS}. For this reason, this low-energy region will not be our focus. 
To identify the unique features of orbital excitations in the block OSMP ($U/W=0.8$), we also present results for the PM ($U/W=0.04$) and FM ($U/W=4$) competing phases.
Overall, we find a close agreement between our dynamical spectra in the block OSMP at $q \simeq 0$ and the two-peak structure observed in experiments, 
involving localized $dd$ excitations of the iron $3d$ orbitals. The OSMP two-peak structure 
(Fig.~\ref{fig:4b}) is distinct from that in the FM phase, where the two peaks are almost identical in position and width. 
With regards to the PM phase results (Figs.~\ref{fig:4a}), there are some similarities with
the spectra shape of the OSMP phase.  The presence of these similarities are natural since
according to Ref.~\cite{JulianOSMP}, at $U/W = 0.04$ the population of orbital $c$ is already near 1, although 
the block magnetic phase is still not fully developed.  
One might argue that the PM regime data could also fit the experiments with appropriate broadening, 
particularly considering that high-frequency results involving interorbital excitations do not typically change abruptly across phase transitions; however, neutron scattering experiments tell us that the material {\it is} magnetically ordered in a block state. Thus the
PM state {\it cannot} be chosen as the correct state even though the RIXS data might be consistent with our predicted dynamical spectra. 

Our model suggests that the experimentally observed peaks represent excitations between localized $d_{xy}$ and itinerant $d_{xz/yz}$ orbitals. 
They are gapped with an activation energy $0.7-0.9$~eV in the calculations, 
and $\sim 0.35-0.45$~eV in the experiments (transitions
involving $e_g$ orbitals, outside our model, should appear at higher energies). 
The difference in absolute gap numbers could be 
fixed by tuning our model parameters still within the OSMP phase 
(or by using ladders instead of chains, which is technically much harder, as discussed before). Since ``fine tuning'' is not our goal,
but a qualitative understanding of results, we consider our (already costly) simulation results sufficient for our main
qualitative conclusion: the block OSMP is the one that most closely resembles the experimental data. 
In addition, in the real samples, there is broadening due to phonons and non-crystallinity effects 
that could make it challenging to define the magnitude of the gap accurately. Regardless, it is clear from experiments 
that there is a gap in the RIXS $q=0$ response. 
However, the paramagnetic phase has a {\it gapless} $q=0$ 
response in $L^{x/y}(q \simeq 0,\w)$ (Fig.~\ref{fig:4a}), 
even-though there is a precursor to a two-peak structure, establishing another difference with the OSMP results. 
Moreover, while the ferromagnetic 
insulator shows a gapped $L^{x/y}(q \simeq 0,\w)$ response, this is with the presence of a {\it single peak} 
at $\w \simeq 1$ eV. In summary, the gapped two-peak response is only evident in the block OSMP phase.


{\bf Summary and Conclusions} --- {\it Our results suggest 
that the experimentally observed gap is not a conventional semi-conducting gap, but instead originates 
from the inter-orbital excitations of a magnetically ordered OSMP}. Figure~\ref{fig:4b} shows that 
peak $A$ occurs at approximately $0.9$~eV, for the zero momentum transfer, as a result of
vertical ($\Delta q = 0$) scattering across $E_F$ from the itinerant $d_{xz/yz}$ 
orbitals to the $d_{xy}$ Mott orbital (Fig.~\ref{fig:3b}, $P_2 \rightarrow P_4$). The 
shoulder/peak labeled as $B$ represents scattering from the localized 
$d_{xy}$ band below $E_F$ to the itinerant hole pocket bands $d_{xz/yz}$ above 
$E_F$, with zero momentum transfer $q/\pi = 0$ (Fig.~\ref{fig:3b}, $P_1 \rightarrow P_3$).
These peaks cannot be described using a simple weak-coupling framework. 


The orbital $d_{xy}$ charge excitations in our calculations 
generate a response at $\w = 1.4$~eV of intensity $\sim 100$ times
smaller than the orbital excitations (Figs.~\ref{fig:2f}-\ref{fig:2i}). This is
because local charge fluctuations are suppressed 
significantly in the Mott orbital, namely the probabilities associated with a doubly occupied 
or empty site configuration are small compared to configurations where sites are half-filled.
Additionally, the features in $\Nkw{a/b}$ and $L^z(q,\w)$ originate  
from itinerant carriers that are sensitive to the incident x-ray energy ($\hbar \omega_\mathrm{in}$) of the RIXS 
experiments. It is known that localized excitations do not shift with $\hbar \omega_\mathrm{in}$, 
while itinerant carriers produce a response that shifts linearly with $\hbar \omega_\mathrm{in}$, 
becoming part of the fluorescence at large $\hbar \omega_\mathrm{in}$ \cite{NicklateRIXS}. 
Therefore, the $\Lkw{z}$ peak at the low-energy transfer $\w$ (Fig.~\ref{fig:4b}, dashed) will shift with
the incident energy. The same is true for itinerant charge excitations $\Nkw{a/b}$. 
In fact, RIXS experiment on BaFe$_2$Se$_3$ also find (fluorescence) 
peaks that shift with incident energy and 
merge with the localized excitations (A and B of Fig.~\ref{fig:4b}), suggesting the existence 
of both localized and itinerant degrees of freedom at the experimental low temperature, as also found
for the OSMP regime in our calculations.

To understand better the characteristics of the states 
at peak positions $A$ and $B$, we also show  
real-space $L_{x/y}$ orbital correlations vs energy transfer at fixed 
distances ($1^{st}$ to $4^{th}$ nearest neighbor) in Fig.~\ref{fig:5}. 
In the block OSMP the high energy states at peak $A$ have positive $L_{x/y}$ correlations up to the $4^{th}$ nearest neighbor in real-space, while states at peak $B$ have negative $L_{x/y}$ correlations only up to the nearest neighbor. {\it In Fig.~\ref{fig:5}, we explicitly show that the peak $A$ of the block OSMP represents states with ferro-orbital ordering, 
while $B$ represents states with short-range anti-ferro-orbital fluctuations}. 
However, the ground state of block OSMP does not have an orbital ordering unlike the ground-state of the competing
ferromagnetic insulator at $U/W = 4$, which has uniform magnetic \cite{JulianOSMP} and 
staggered $L_z$ orbital order (Fig.~\ref{fig:2h}). 
For the $U/W = 4$ FM case, the corresponding $L_{x/y}$ is gapped and has excitations 
only at $\w \sim 1.0$ (Fig.~\ref{fig:2h}). In real-space, these high-energy states 
show short-range positive correlations, persisting only up to the $2^{nd}$ nearest neighbor (not shown).


%
\begin{figure}[t]
\begin{center}
\begin{overpic}[trim = 0cm 0cm 0cm 0cm,angle=0]{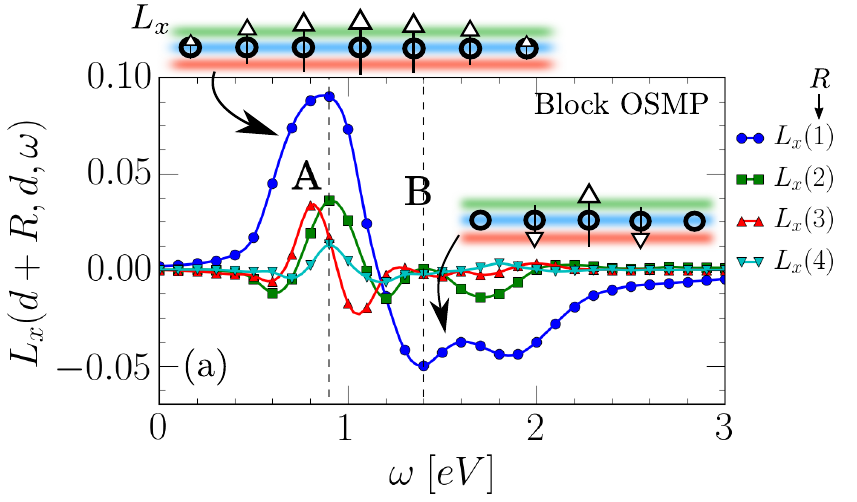}
\end{overpic}
\end{center}
\subfloat{\label{fig:5}}
\vspace{-1.0cm}
\caption{Real-space $L_x(d+R,d,\w)$ [using $O_i = L^{x}_{d+R}$ and $O_j = L^{x}_{d}$ 
in Eq.(1)] vs. $\omega$ with a
fixed center site $d = 8$ on a $16$-sites three-orbital chain in the block OSMP ($U/W=0.8$). 
Labels $A$ and $B$ (vertical dashed line) show the peak positions labels 
in Fig.~\ref{fig:4}. 
Overall negative (positive) response denote local antiferro (ferro) orbital ordering. 
The cartoons represent pictorially the $L_{x/y}$ correlations described in the text that
characterize the A and B states. 
}
\label{fig:5}
\end{figure}
%

{\bf  Conclusions} ---
We calculated the momentum-resolved orbital and charge dynamics 
of an orbital-selective Mott phase using the three $t_{2g}$ orbitals, 
and compared our results to the available RIXS data for 
BaFe$_2$Se$_3$ at zero momentum transfer \cite{MonneyRIXS}. 
We find localized $dd$ excitations that produce particular peaks $A$ and $B$ that are 
very similar 
to those observed experimentally, providing theoretical support that BaFe$_2$Se$_3$ is indeed an orbital-selective Mott insulator. 
Moreover, we predict the
$q$-resolved charge and orbital dynamical spectra that can be measured by RIXS in the future. 
Recent RIXS experiments display a very similar response 
for BaFe$_2$Se$_3$ and BaFe$_2$S$_3$, implying that an orbital-selective phase is present 
in the  BaFe$_2$S$_3$ high-pressure superconductor as well \cite{NatMatSC,RIXSBaFeX}. We encourage the measurement of the band structure and density-of-states of these materials using angle-resolved photoemission and scanning tunneling microscopy, to 
confirm our predicted band(s) crossing of the Fermi surface and Mott like band(s) below the 
Fermi surface. We also encourage future $q$-resolved RIXS measurements 
for BaFe$_2$Se$_3$ and BaFe$_2$S$_3$ compounds. 
Finally, the novel orbital fractionalization proposed in the FM insulator phase 
defines a new avenue of research that will be pursued soon.

\vskip 1 cm
\noindent{\bf Methods}
\newline
{\bf Model} ---  
We use a multi-orbital Hubbard model composed of kinetic energy 
and interaction terms as $H = H_{K} + H_{I}$. The kinetic energy part of the 
Hamiltonian is 
%
\begin{equation*}\label{eq:HKmodel}
\begin{split}
H_{K} &= \sum_{i,\sigma,\gamma,\gamma^\prime} t^{\gamma \gamma^\prime} 
( \cd{i \gamma \sigma} \c{i+1 \gamma^\prime \sigma}^\pdag + \text{H.c.} ) + \sum_{i,\sigma,\gamma} \epsilon_{\gamma} n_{i \sigma \gamma},
\end{split}
\end{equation*}
%
where $\cd{i \gamma \sigma}$ ($\c{i \gamma \sigma}$) creates (destroys) 
an electron at site $i$, orbital $\gamma$, 
and spin $\sigma$. The first term represents nearest-neighbor electron hopping 
from orbital $\gamma'$ to $\gamma$ with a hopping amplitude 
$t^{\gamma \gamma'}$. We denote the relevant 
orbitals $d_{xz}$, $d_{yz}$, and $d_{xy}$ as $a$, $b$, and $c$, respectively. 
The second term contains the orbital-dependent crystal field splitting. 
The parameters are (eV units) $\epsilon_a =-0.1, \epsilon_b=0.0$,  $\epsilon_c=0.8$, 
$t^{aa} = t^{bb} = 0.5$, $t^{cc} = 0.15$, and $t^{ac} = t^{bc} = t^{ca} = t^{cb} = -0.10$.
The non-interacting bandwidth is $W = 4.9t^{aa}$. 
This set of parameters is known \cite{JulianOSMP} to produce bands that emulate 
iron-based superconductors, with hole pockets 
at $q = 0$ and an electron pocket at $q=\pm \pi$ \cite{nphys2438}. Our reported results are all at zero temperature.

The interaction term, in standard notation, is 
%
\begin{equation*}\label{eq:HImodel}
\begin{split}
H_{I} &= U \sum_{i, \gamma} n_{i \gamma \uparrow} n_{i \gamma \downarrow} 
		+ (U^\prime- \frac{J_{H}}{2}) \sum_{\substack{i, \gamma < \gamma'}} n_{i \gamma} n_{i \gamma'} \\
     &- 2J_{H} \sum_{\substack{i, \gamma < \gamma'}} {{\mathbf{S}_{i \gamma}}\cdot{\mathbf{S}_{i \gamma'}}} 
     + J_{H} \sum_{\substack{i, \gamma < \gamma'}} ( P^{\dagger}_{i\gamma}P^\pdag_{i\gamma'} + \text{H.c.}).
\end{split}
\end{equation*}
%
The first term is the intraorbital Hubbard repulsion $U$. 
The second is the interorbital repulsion between electrons at different orbitals, with $U^\prime = U - 2J_H$. 
The third term is the Hund's coupling $J_H$, 
and the last term represents the on-site interorbital hopping of electron pairs
($P_{i \gamma'} = c_{i \gamma' \uparrow} c_{i \gamma' \downarrow}$). 
Explicit definitions for all the operators in the model are in 
the supplementary material sections I and II.
The rich physical properties realized by this model were discussed extensively in
Ref.~\cite{JulianOSMP}. 
Most of the data presented here was gathered at $J_H/U = 0.25$ 
(as used extensively before~\cite{haule,luo2010,JulianOSMP}) and
$U/W = 0.8$ where the block-type AFM OSMP is known to be stable \cite{JulianOSMP}. 
However, we also show results for the 
paramagnetic metal (PM, small $U$) and ferromagnetic insulator (FM, large $U$) 
phases to highlight unique features of the block OSMP by contrast.

{\bf Observables} --- To characterize the OSMP, we calculate the 
dynamical response functions
%
\begin{equation} \label{eq:Green}
\begin{split}
O(i,j,\omega) &= \frac{-1}{\pi} \text{Im} \big[ \langle \psi_0 | O^{\dagger}_{i} \frac{1}{\omega - H + E_g + {\mathrm i}\eta} O_{j} | \psi_0 \rangle \big]
\end{split} 
\end{equation}
%
using DMRG within the correction-vector formulation in Krylov 
space \cite{NoceraKrylov,WhiteCorrVec}.
The single-particle photoemission spectral function is obtained using 
$O_i=\c{i \sigma \gamma}$. 
The intraorbital particle-hole (charge) excitations arise using 
$O_i = \sum_{\sigma} n_{i \sigma \gamma} - \< n_{i \sigma \gamma} \>$, 
where we explicitly subtract the ground-state contribution to measure only the fluctuations. 
Finally, the orbital excitations are obtained using $O_i = \{L^x_i, L^y_i, L^z_i\}$, where 
%
\begin{equation} \label{eq:Green}
\begin{split} 
L^{x}_{i} = {\mathrm i} \sum_\sigma (\cd{ia\sigma} \c{ic\sigma}^\pdag - \cd{ic\sigma} \c{ia\sigma}^\pdag), \\
L^{y}_{i} = {\mathrm i} \sum_\sigma (\cd{ib\sigma} \c{ic\sigma}^\pdag - \cd{ic\sigma} \c{ib\sigma}^\pdag), \\
L^{z}_{i} = {\mathrm i} \sum_\sigma (\cd{ia\sigma} \c{ib\sigma}^\pdag - \cd{ib\sigma} \c{ia\sigma}^\pdag).
\end{split} 
\end{equation}
%
These operators are derived from the $L=2$ angular momentum operators 
in the $t_{2g}$ orbital basis, see supplemental material section II.


{\bf  Code Availability} --- 
Computer codes used in this study are available at \url{https://g1257.github.io/dmrgPlusPlus/}. \\

{\bf  Data Availability} --- 
The data that support the findings of this study are available from the corresponding
author upon request. \\

\bibliographystyle{naturemag}
\bibliography{citations}

\begin{thebibliography}{10}
\expandafter\ifx\csname url\endcsname\relax
  \def\url#1{\texttt{#1}}\fi
\expandafter\ifx\csname urlprefix\endcsname\relax\def\urlprefix{URL }\fi
\providecommand{\bibinfo}[2]{#2}
\providecommand{\eprint}[2][]{\url{#2}}

\bibitem{NatMatSC}
\bibinfo{author}{Takahashi, H.} \emph{et~al.}
\newblock \bibinfo{title}{Pressure-induced superconductivity in the
  {iron}-based ladder material {${\mathrm{BaFe}}_{2}{\mathrm{S}}_{3}$}}.
\newblock \emph{\bibinfo{journal}{Nature Materials}}
  \textbf{\bibinfo{volume}{14}}, \bibinfo{pages}{1008} (\bibinfo{year}{2015}).
\newblock \urlprefix\url{http://dx.doi.org/10.1038/nmat4351}.

\bibitem{PRLSC}
\bibinfo{author}{Yamauchi, T.}, \bibinfo{author}{Hirata, Y.},
  \bibinfo{author}{Ueda, Y.} \& \bibinfo{author}{Ohgushi, K.}
\newblock \bibinfo{title}{Pressure-induced {Mott} transition followed by a
  {24-K} superconducting phase in {${\mathrm{BaFe}}_{2}{\mathrm{S}}_{3}$}}.
\newblock \emph{\bibinfo{journal}{Phys. Rev. Lett.}}
  \textbf{\bibinfo{volume}{115}}, \bibinfo{pages}{246402}
  (\bibinfo{year}{2015}).
\newblock
  \urlprefix\url{https://link.aps.org/doi/10.1103/PhysRevLett.115.246402}.

\bibitem{123SeSC1}
\bibinfo{author}{Ying, J.}, \bibinfo{author}{Lei, H.},
  \bibinfo{author}{Petrovic, C.}, \bibinfo{author}{Xiao, Y.} \&
  \bibinfo{author}{Struzhkin, V.~V.}
\newblock \bibinfo{title}{Interplay of magnetism and superconductivity in the
  compressed {Fe}-ladder compound {${\mathrm{BaFe}}_{2}{\mathrm{Se}}_{3}$}}.
\newblock \emph{\bibinfo{journal}{Phys. Rev. B}} \textbf{\bibinfo{volume}{95}},
  \bibinfo{pages}{241109} (\bibinfo{year}{2017}).
\newblock \urlprefix\url{https://link.aps.org/doi/10.1103/PhysRevB.95.241109}.

\bibitem{123SeSC2}
\bibinfo{author}{Zhang, Y.}, \bibinfo{author}{Lin, L.-F.},
  \bibinfo{author}{Zhang, J.-J.}, \bibinfo{author}{Dagotto, E.} \&
  \bibinfo{author}{Dong, S.}
\newblock \bibinfo{title}{Sequential structural and antiferromagnetic
  transitions in {${\mathrm{BaFe}}_{2}{\mathrm{Se}}_{3}$} under pressure}.
\newblock \emph{\bibinfo{journal}{Phys. Rev. B}} \textbf{\bibinfo{volume}{97}},
  \bibinfo{pages}{045119} (\bibinfo{year}{2018}).
\newblock \urlprefix\url{https://link.aps.org/doi/10.1103/PhysRevB.97.045119}.

\bibitem{Pottgen1}
\bibinfo{author}{Svitlyk, V.} \emph{et~al.}
\newblock \bibinfo{title}{Crystal structure of {BaFe$_2$Se$_3$} as a function
  of temperature and pressure: phase transition phenomena and high-order
  expansion of {Landau} potential}.
\newblock \emph{\bibinfo{journal}{Journal of Physics: Condensed Matter}}
  \textbf{\bibinfo{volume}{25}}, \bibinfo{pages}{315403}
  (\bibinfo{year}{2013}).
\newblock \urlprefix\url{http://stacks.iop.org/0953-8984/25/i=31/a=315403}.

\bibitem{ironSCfamily1}
\bibinfo{author}{Basov, D.} \& \bibinfo{author}{Chubukov, A.~V.}
\newblock \bibinfo{title}{Manifesto for a higher {$T_c$}}.
\newblock \emph{\bibinfo{journal}{Nature Physics}}
  \textbf{\bibinfo{volume}{7}}, \bibinfo{pages}{272} (\bibinfo{year}{2011}).
\newblock \urlprefix\url{https://www.nature.com/articles/nphys1975}.

\bibitem{ironSCfamily2}
\bibinfo{author}{Fernandes, R.~M.} \& \bibinfo{author}{Chubukov, A.~V.}
\newblock \bibinfo{title}{Low-energy microscopic models for {iron}-based
  superconductors: a review}.
\newblock \emph{\bibinfo{journal}{Reports on Progress in Physics}}
  \textbf{\bibinfo{volume}{80}}, \bibinfo{pages}{014503}
  (\bibinfo{year}{2017}).
\newblock \urlprefix\url{http://stacks.iop.org/0034-4885/80/i=1/a=014503}.

\bibitem{nphys2438}
\bibinfo{author}{Dai, P.}, \bibinfo{author}{Hu, J.} \&
  \bibinfo{author}{Dagotto, E.}
\newblock \bibinfo{title}{Magnetism and its microscopic origin in {iron}-based
  high-temperature superconductors}.
\newblock \emph{\bibinfo{journal}{Nature Physics}}
  \textbf{\bibinfo{volume}{85}}, \bibinfo{pages}{709--718}
  (\bibinfo{year}{2012}).
\newblock \urlprefix\url{http://dx.doi.org/10.1038/nphys2438}.

\bibitem{DagottoRMP13}
\bibinfo{author}{Dagotto, E.}
\newblock \bibinfo{title}{Colloquium: The unexpected properties of alkali metal
  {iron} selenide superconductors}.
\newblock \emph{\bibinfo{journal}{Rev. Mod. Phys.}}
  \textbf{\bibinfo{volume}{85}}, \bibinfo{pages}{849--867}
  (\bibinfo{year}{2013}).
\newblock \urlprefix\url{https://link.aps.org/doi/10.1103/RevModPhys.85.849}.

\bibitem{WhiteDMRG1}
\bibinfo{author}{White, S.~R.}
\newblock \bibinfo{title}{Density matrix formulation for quantum
  renormalization groups}.
\newblock \emph{\bibinfo{journal}{Phys. Rev. Lett.}}
  \textbf{\bibinfo{volume}{69}}, \bibinfo{pages}{2863--2866}
  (\bibinfo{year}{1992}).
\newblock \urlprefix\url{https://link.aps.org/doi/10.1103/PhysRevLett.69.2863}.

\bibitem{DMRGreview}
\bibinfo{author}{Schollw\"ock, U.}
\newblock \bibinfo{title}{The density-matrix renormalization group}.
\newblock \emph{\bibinfo{journal}{Rev. Mod. Phys.}}
  \textbf{\bibinfo{volume}{77}}, \bibinfo{pages}{259--315}
  (\bibinfo{year}{2005}).
\newblock \urlprefix\url{https://link.aps.org/doi/10.1103/RevModPhys.77.259}.

\bibitem{white2005}
\bibinfo{author}{White, S.~R.}
\newblock \bibinfo{title}{Density matrix renormalization group algorithms with
  a single center site}.
\newblock \emph{\bibinfo{journal}{Phys. Rev. B}} \textbf{\bibinfo{volume}{72}},
  \bibinfo{pages}{180403} (\bibinfo{year}{2005}).
\newblock \urlprefix\url{https://link.aps.org/doi/10.1103/PhysRevB.72.180403}.

\bibitem{WhiteWFT}
\bibinfo{author}{White, S.~R.}
\newblock \bibinfo{title}{Spin gaps in a frustrated {Heisenberg} model for
  {${\mathrm{CaV}}_{4}{O}_{9}$}}.
\newblock \emph{\bibinfo{journal}{Phys. Rev. Lett.}}
  \textbf{\bibinfo{volume}{77}}, \bibinfo{pages}{3633--3636}
  (\bibinfo{year}{1996}).
\newblock \urlprefix\url{https://link.aps.org/doi/10.1103/PhysRevLett.77.3633}.

\bibitem{GonzaloDMRGpp}
\bibinfo{author}{Alvarez, G.}
\newblock \bibinfo{title}{The density matrix renormalization group for strongly
  correlated electron systems: A generic implementation}.
\newblock \emph{\bibinfo{journal}{Computer Physics Communications}}
  \textbf{\bibinfo{volume}{180}}, \bibinfo{pages}{1572--1578}
  (\bibinfo{year}{2009}).
\newblock \urlprefix\url{https://doi.org/10.1016/j.cpc.2009.02.016}.

\bibitem{PatelBaFe2S3}
\bibinfo{author}{Patel, N.~D.} \emph{et~al.}
\newblock \bibinfo{title}{Magnetic properties and pairing tendencies of the
  {iron}-based superconducting ladder {${\mathrm{BaFe}}_{2}{\mathrm{S}}_{3}$}:
  Combined ab initio and density matrix renormalization group study}.
\newblock \emph{\bibinfo{journal}{Phys. Rev. B}} \textbf{\bibinfo{volume}{94}},
  \bibinfo{pages}{075119} (\bibinfo{year}{2016}).
\newblock \urlprefix\url{https://link.aps.org/doi/10.1103/PhysRevB.94.075119}.

\bibitem{NambuBaFeSe}
\bibinfo{author}{Nambu, Y.} \emph{et~al.}
\newblock \bibinfo{title}{Block magnetism coupled with local distortion in the
  {iron}-based spin-ladder compound {BaFe${}_{2}$Se${}_{3}$}}.
\newblock \emph{\bibinfo{journal}{Phys. Rev. B}} \textbf{\bibinfo{volume}{85}},
  \bibinfo{pages}{064413} (\bibinfo{year}{2012}).
\newblock \urlprefix\url{https://link.aps.org/doi/10.1103/PhysRevB.85.064413}.

\bibitem{Patel2OrbChain}
\bibinfo{author}{Patel, N.~D.}, \bibinfo{author}{Nocera, A.},
  \bibinfo{author}{Alvarez, G.}, \bibinfo{author}{Moreo, A.} \&
  \bibinfo{author}{Dagotto, E.}
\newblock \bibinfo{title}{Pairing tendencies in a two-orbital {Hubbard} model
  in one dimension}.
\newblock \emph{\bibinfo{journal}{Phys. Rev. B}} \textbf{\bibinfo{volume}{96}},
  \bibinfo{pages}{024520} (\bibinfo{year}{2017}).
\newblock \urlprefix\url{https://link.aps.org/doi/10.1103/PhysRevB.96.024520}.

\bibitem{CaronBaFeSe}
\bibinfo{author}{Caron, J.~M.}, \bibinfo{author}{Neilson, J.~R.},
  \bibinfo{author}{Miller, D.~C.}, \bibinfo{author}{Llobet, A.} \&
  \bibinfo{author}{McQueen, T.~M.}
\newblock \bibinfo{title}{{Iron} displacements and magnetoelastic coupling in
  the antiferromagnetic spin-ladder compound {BaFe${}_{2}$Se${}_{3}$}}.
\newblock \emph{\bibinfo{journal}{Phys. Rev. B}} \textbf{\bibinfo{volume}{84}},
  \bibinfo{pages}{180409} (\bibinfo{year}{2011}).
\newblock \urlprefix\url{https://link.aps.org/doi/10.1103/PhysRevB.84.180409}.

\bibitem{FeLadder2}
\bibinfo{author}{Lei, H.}, \bibinfo{author}{Ryu, H.}, \bibinfo{author}{Frenkel,
  A.~I.} \& \bibinfo{author}{Petrovic, C.}
\newblock \bibinfo{title}{Anisotropy in {BaFe${}_{2}$Se${}_{3}$} single
  crystals with double chains of {FeSe} tetrahedra}.
\newblock \emph{\bibinfo{journal}{Phys. Rev. B}} \textbf{\bibinfo{volume}{84}},
  \bibinfo{pages}{214511} (\bibinfo{year}{2011}).
\newblock \urlprefix\url{https://link.aps.org/doi/10.1103/PhysRevB.84.214511}.

\bibitem{FeLadder5}
\bibinfo{author}{Caron, J.~M.} \emph{et~al.}
\newblock \bibinfo{title}{Orbital-selective magnetism in the spin-ladder {iron}
  selenides {Ba${}_{1\ensuremath{-}x}$K${}_{x}$Fe${}_{2}$Se${}_{3}$}}.
\newblock \emph{\bibinfo{journal}{Phys. Rev. B}} \textbf{\bibinfo{volume}{85}},
  \bibinfo{pages}{180405} (\bibinfo{year}{2012}).
\newblock \urlprefix\url{https://link.aps.org/doi/10.1103/PhysRevB.85.180405}.

\bibitem{FeLadder7}
\bibinfo{author}{Luo, Q.} \emph{et~al.}
\newblock \bibinfo{title}{Magnetic states of the two-leg-ladder alkali metal
  {iron} selenides {$A$Fe${}_{2}$Se${}_{3}$}}.
\newblock \emph{\bibinfo{journal}{Phys. Rev. B}} \textbf{\bibinfo{volume}{87}},
  \bibinfo{pages}{024404} (\bibinfo{year}{2013}).
\newblock \urlprefix\url{https://link.aps.org/doi/10.1103/PhysRevB.87.024404}.

\bibitem{NeutronOSMP}
\bibinfo{author}{Mourigal, M.} \emph{et~al.}
\newblock \bibinfo{title}{Block magnetic excitations in the orbitally selective
  {Mott} insulator {${\mathrm{BaFe}}_{2}{\mathrm{Se}}_{3}$}}.
\newblock \emph{\bibinfo{journal}{Phys. Rev. Lett.}}
  \textbf{\bibinfo{volume}{115}}, \bibinfo{pages}{047401}
  (\bibinfo{year}{2015}).
\newblock
  \urlprefix\url{https://link.aps.org/doi/10.1103/PhysRevLett.115.047401}.

\bibitem{MonneyRIXS}
\bibinfo{author}{Monney, C.} \emph{et~al.}
\newblock \bibinfo{title}{Resonant inelastic {X}-ray scattering at the {Fe}
  {${L}_{3}$} edge of the one-dimensional chalcogenide
  {BaFe${}_{2}$Se${}_{3}$}}.
\newblock \emph{\bibinfo{journal}{Phys. Rev. B}} \textbf{\bibinfo{volume}{88}},
  \bibinfo{pages}{165103} (\bibinfo{year}{2013}).
\newblock \urlprefix\url{https://link.aps.org/doi/10.1103/PhysRevB.88.165103}.

\bibitem{PEOhgushi}
\bibinfo{author}{Ootsuki, D.} \emph{et~al.}
\newblock \bibinfo{title}{Coexistence of localized and itinerant electrons in
  {${\mathrm{BaFe}}_{2}{X}_{3}$} {$(X=\mathrm{S}$} and {$\mathrm{Se}$})
  revealed by photoemission spectroscopy}.
\newblock \emph{\bibinfo{journal}{Phys. Rev. B}} \textbf{\bibinfo{volume}{91}},
  \bibinfo{pages}{014505} (\bibinfo{year}{2015}).
\newblock \urlprefix\url{https://link.aps.org/doi/10.1103/PhysRevB.91.014505}.

\bibitem{georges}
\bibinfo{author}{Georges, A.}, \bibinfo{author}{d\'e Medici, L.} \&
  \bibinfo{author}{Mravlje, J.}
\newblock \bibinfo{title}{Strong correlations from {Hund's} coupling}.
\newblock \emph{\bibinfo{journal}{Annual Review of Condensed Matter Physics}}
  \textbf{\bibinfo{volume}{4}}, \bibinfo{pages}{137--178}
  (\bibinfo{year}{2013}).
\newblock
  \urlprefix\url{https://doi.org/10.1146/annurev-conmatphys-020911-125045}.
\newblock \bibinfo{note}{And references therein}.

\bibitem{yu-si}
\bibinfo{author}{Yu, R.} \& \bibinfo{author}{Si, Q.}
\newblock \bibinfo{title}{Orbital-selective {Mott} phase in multiorbital models
  for alkaline {iron} {selenides}
  {${\mathrm{K}}_{1\ensuremath{-}x}{\mathrm{Fe}}_{2\ensuremath{-}y}{\mathrm{Se}}_{2}$}}.
\newblock \emph{\bibinfo{journal}{Phys. Rev. Lett.}}
  \textbf{\bibinfo{volume}{110}}, \bibinfo{pages}{146402}
  (\bibinfo{year}{2013}).
\newblock
  \urlprefix\url{https://link.aps.org/doi/10.1103/PhysRevLett.110.146402}.

\bibitem{JulianOSMP}
\bibinfo{author}{Rinc\'on, J.}, \bibinfo{author}{Moreo, A.},
  \bibinfo{author}{Alvarez, G.} \& \bibinfo{author}{Dagotto, E.}
\newblock \bibinfo{title}{Exotic magnetic order in the orbital-selective {Mott}
  regime of multiorbital systems}.
\newblock \emph{\bibinfo{journal}{Phys. Rev. Lett.}}
  \textbf{\bibinfo{volume}{112}}, \bibinfo{pages}{106405}
  (\bibinfo{year}{2014}).
\newblock
  \urlprefix\url{https://link.aps.org/doi/10.1103/PhysRevLett.112.106405}.

\bibitem{Jacekpreparation}
\bibinfo{note}{Herbrych, J., Heverhagen, J., Patel, N. D., Alvarez, G.,
  Daghofer, M., Moreo, A. {\&} Dagotto, E., in preparation.}

\bibitem{JacekSpinOSMP}
\bibinfo{author}{Herbrych, J.} \emph{et~al.}
\newblock \bibinfo{title}{Spin dynamics of the block orbital-selective {Mott}
  phase}.
\newblock \emph{\bibinfo{journal}{Nature Communications}}
  \textbf{\bibinfo{volume}{9}}, \bibinfo{pages}{3736} (\bibinfo{year}{2018}).
\newblock \urlprefix\url{https://www.nature.com/articles/s41467-018-06181-6}.

\bibitem{SLiOSMP}
\bibinfo{author}{Li, S.} \emph{et~al.}
\newblock \bibinfo{title}{Nonlocal correlations in the orbital selective {Mott}
  phase of a one-dimensional multiorbital {Hubbard} model}.
\newblock \emph{\bibinfo{journal}{Phys. Rev. B}} \textbf{\bibinfo{volume}{94}},
  \bibinfo{pages}{235126} (\bibinfo{year}{2016}).
\newblock \urlprefix\url{https://link.aps.org/doi/10.1103/PhysRevB.94.235126}.

\bibitem{RMPRIXS1}
\bibinfo{author}{Ament, L. J.~P.}, \bibinfo{author}{van Veenendaal, M.},
  \bibinfo{author}{Devereaux, T.~P.}, \bibinfo{author}{Hill, J.~P.} \&
  \bibinfo{author}{van~den Brink, J.}
\newblock \bibinfo{title}{Resonant inelastic {X}-ray scattering studies of
  elementary excitations}.
\newblock \emph{\bibinfo{journal}{Rev. Mod. Phys.}}
  \textbf{\bibinfo{volume}{83}}, \bibinfo{pages}{705--767}
  (\bibinfo{year}{2011}).
\newblock \urlprefix\url{https://link.aps.org/doi/10.1103/RevModPhys.83.705}.

\bibitem{RIXS1}
\bibinfo{author}{Wohlfeld, K.}, \bibinfo{author}{Nishimoto, S.},
  \bibinfo{author}{Haverkort, M.~W.} \& \bibinfo{author}{van~den Brink, J.}
\newblock \bibinfo{title}{Microscopic origin of spin-orbital separation in
  {Sr${}_{2}$CuO${}_{3}$}}.
\newblock \emph{\bibinfo{journal}{Phys. Rev. B}} \textbf{\bibinfo{volume}{88}},
  \bibinfo{pages}{195138} (\bibinfo{year}{2013}).
\newblock \urlprefix\url{https://link.aps.org/doi/10.1103/PhysRevB.88.195138}.

\bibitem{RIXS2}
\bibinfo{author}{Marra, P.}, \bibinfo{author}{Wohlfeld, K.} \&
  \bibinfo{author}{van~den Brink, J.}
\newblock \bibinfo{title}{Unraveling orbital correlations with magnetic
  resonant inelastic {X}-ray scattering}.
\newblock \emph{\bibinfo{journal}{Phys. Rev. Lett.}}
  \textbf{\bibinfo{volume}{109}}, \bibinfo{pages}{117401}
  (\bibinfo{year}{2012}).
\newblock
  \urlprefix\url{https://link.aps.org/doi/10.1103/PhysRevLett.109.117401}.

\bibitem{RIXS3}
\bibinfo{author}{Jia, C.} \emph{et~al.}
\newblock \bibinfo{title}{Persistent spin excitations in doped antiferromagnets
  revealed by resonant inelastic light scattering}.
\newblock \emph{\bibinfo{journal}{Nature Communications}}
  \textbf{\bibinfo{volume}{5}}, \bibinfo{pages}{3314} (\bibinfo{year}{2014}).
\newblock \urlprefix\url{https://www.nature.com/articles/ncomms4314}.

\bibitem{JohnstonNatureComm}
\bibinfo{author}{Johnston, S.} \emph{et~al.}
\newblock \bibinfo{title}{Electron-lattice interactions strongly renormalize
  the charge-transfer energy in the spin-chain cuprate {Li$_2$CuO$_2$}}.
\newblock \emph{\bibinfo{journal}{Nature Communications}}
  \textbf{\bibinfo{volume}{7}}, \bibinfo{pages}{10563} (\bibinfo{year}{2016}).
\newblock \urlprefix\url{https://www.nature.com/articles/ncomms10563}.

\bibitem{rincon2}
\bibinfo{note}{For an analysis of the full phase diagram varying the electronic
  density, and for a discussion of several {OSMP} states see Rinc\'on, J. {\it
  et al.} Quantum phase transition between orbital-selective {Mott} states in
  {Hund's} metals {\it Phys. Rev. B} {\bf 90}, 241105 (2014). URL
  \url{https://link.aps.org/doi/10.1103/PhysRevB.90.241105}}.

\bibitem{footnote-cgap}
\bibinfo{note}{Note that the finite weight at $\mu$ in the $c$ orbital is due
  to broadening.}

\bibitem{AlbertoRPAvsFLEX}
\bibinfo{author}{Nocera, A.} \emph{et~al.}
\newblock \bibinfo{title}{Doping evolution of charge and spin excitations in
  two-leg {Hubbard} ladders: Comparing {DMRG} and {FLEX} results}.
\newblock \emph{\bibinfo{journal}{Phys. Rev. B}} \textbf{\bibinfo{volume}{97}},
  \bibinfo{pages}{195156} (\bibinfo{year}{2018}).
\newblock \urlprefix\url{https://link.aps.org/doi/10.1103/PhysRevB.97.195156}.

\bibitem{schlappa2012spin}
\bibinfo{author}{Schlappa, J.} \emph{et~al.}
\newblock \bibinfo{title}{Spin-orbital separation in the quasi-one-dimensional
  {Mott} insulator {${\mathrm{Sr}}_{2}{\mathrm{CuO}}_{3}$}}.
\newblock \emph{\bibinfo{journal}{Nature}} \textbf{\bibinfo{volume}{485}},
  \bibinfo{pages}{82} (\bibinfo{year}{2012}).
\newblock \urlprefix\url{https://www.nature.com/articles/nature10974}.

\bibitem{orbext1}
\bibinfo{author}{Chen, C.-C.}, \bibinfo{author}{van Veenendaal, M.},
  \bibinfo{author}{Devereaux, T.~P.} \& \bibinfo{author}{Wohlfeld, K.}
\newblock \bibinfo{title}{Fractionalization, entanglement, and separation:
  Understanding the collective excitations in a spin-orbital chain}.
\newblock \emph{\bibinfo{journal}{Phys. Rev. B}} \textbf{\bibinfo{volume}{91}},
  \bibinfo{pages}{165102} (\bibinfo{year}{2015}).
\newblock \urlprefix\url{https://link.aps.org/doi/10.1103/PhysRevB.91.165102}.

\bibitem{orbext2}
\bibinfo{author}{Wohlfeld, K.}, \bibinfo{author}{Daghofer, M.},
  \bibinfo{author}{Nishimoto, S.}, \bibinfo{author}{Khaliullin, G.} \&
  \bibinfo{author}{van~den Brink, J.}
\newblock \bibinfo{title}{Intrinsic coupling of orbital excitations to spin
  fluctuations in mott insulators}.
\newblock \emph{\bibinfo{journal}{Phys. Rev. Lett.}}
  \textbf{\bibinfo{volume}{107}}, \bibinfo{pages}{147201}
  (\bibinfo{year}{2011}).
\newblock
  \urlprefix\url{https://link.aps.org/doi/10.1103/PhysRevLett.107.147201}.

\bibitem{orbext3}
\bibinfo{author}{Heverhagen, J.} \& \bibinfo{author}{Daghofer, M.}
\newblock \bibinfo{title}{Spinon-orbiton repulsion and attraction mediated by
  {Hund's} rule}.
\newblock \emph{\bibinfo{journal}{Phys. Rev. B}} \textbf{\bibinfo{volume}{98}},
  \bibinfo{pages}{085120} (\bibinfo{year}{2018}).
\newblock \urlprefix\url{https://link.aps.org/doi/10.1103/PhysRevB.98.085120}.

\bibitem{JiaPRX}
\bibinfo{author}{Jia, C.}, \bibinfo{author}{Wohlfeld, K.},
  \bibinfo{author}{Wang, Y.}, \bibinfo{author}{Moritz, B.} \&
  \bibinfo{author}{Devereaux, T.~P.}
\newblock \bibinfo{title}{Using {RIXS} to uncover elementary charge and spin
  excitations}.
\newblock \emph{\bibinfo{journal}{Phys. Rev. X}} \textbf{\bibinfo{volume}{6}},
  \bibinfo{pages}{021020} (\bibinfo{year}{2016}).
\newblock \urlprefix\url{https://link.aps.org/doi/10.1103/PhysRevX.6.021020}.

\bibitem{Monneyprivate}
\bibinfo{note}{Monney, C., private communication.}

\bibitem{NicklateRIXS}
\bibinfo{author}{Bisogni, V.} \emph{et~al.}
\newblock \bibinfo{title}{Ground-state oxygen holes and the metal-insulator
  transition in the negative charge-transfer rare-earth nickelates}.
\newblock \emph{\bibinfo{journal}{Nature Communications}}
  \textbf{\bibinfo{volume}{7}}, \bibinfo{pages}{13017} (\bibinfo{year}{2016}).
\newblock \urlprefix\url{http://dx.doi.org/10.1038/ncomms13017}.

\bibitem{RIXSBaFeX}
\bibinfo{author}{Takubo, K.} \emph{et~al.}
\newblock \bibinfo{title}{Orbital order and fluctuations in the two-leg ladder
  materials {${\mathrm{BaFe}}_{2}{X}_{3}$} ({$X=\mathrm{S}$} and
  {$\mathrm{Se}$}) and {${\mathrm{CsFe}}_{2}{\mathrm{Se}}_{3}$}}.
\newblock \emph{\bibinfo{journal}{Phys. Rev. B}} \textbf{\bibinfo{volume}{96}},
  \bibinfo{pages}{115157} (\bibinfo{year}{2017}).
\newblock \urlprefix\url{https://link.aps.org/doi/10.1103/PhysRevB.96.115157}.

\bibitem{haule}
\bibinfo{author}{Haule, K.} \& \bibinfo{author}{Kotliar, G.}
\newblock \bibinfo{title}{Coherence-incoherence crossover in the normal state
  of iron oxypnictides and importance of {Hund's} rule coupling}.
\newblock \emph{\bibinfo{journal}{New Journal of Physics}}
  \textbf{\bibinfo{volume}{11}}, \bibinfo{pages}{025021}
  (\bibinfo{year}{2009}).
\newblock \urlprefix\url{http://stacks.iop.org/1367-2630/11/i=2/a=025021}.

\bibitem{luo2010}
\bibinfo{author}{Luo, Q.} \emph{et~al.}
\newblock \bibinfo{title}{Neutron and {ARPES} constraints on the couplings of
  the multiorbital {Hubbard} model for the iron pnictides}.
\newblock \emph{\bibinfo{journal}{Phys. Rev. B}} \textbf{\bibinfo{volume}{82}},
  \bibinfo{pages}{104508} (\bibinfo{year}{2010}).
\newblock \urlprefix\url{https://link.aps.org/doi/10.1103/PhysRevB.82.104508}.

\bibitem{NoceraKrylov}
\bibinfo{author}{Nocera, A.} \& \bibinfo{author}{Alvarez, G.}
\newblock \bibinfo{title}{Spectral functions with the density matrix
  renormalization group: {Krylov}-space approach for correction vectors}.
\newblock \emph{\bibinfo{journal}{Phys. Rev. E}} \textbf{\bibinfo{volume}{94}},
  \bibinfo{pages}{053308} (\bibinfo{year}{2016}).
\newblock \urlprefix\url{https://link.aps.org/doi/10.1103/PhysRevE.94.053308}.

\bibitem{WhiteCorrVec}
\bibinfo{author}{K\"uhner, T.~D.} \& \bibinfo{author}{White, S.~R.}
\newblock \bibinfo{title}{Dynamical correlation functions using the density
  matrix renormalization group}.
\newblock \emph{\bibinfo{journal}{Phys. Rev. B}} \textbf{\bibinfo{volume}{60}},
  \bibinfo{pages}{335--343} (\bibinfo{year}{1999}).
\newblock \urlprefix\url{https://link.aps.org/doi/10.1103/PhysRevB.60.335}.

\end{thebibliography}

{\bf  Acknowledgments} --- 
We thank C. Monney and T. Schmitt for providing a copy of the
experimental data shown in Fig.~\ref{fig:4}.
N.~D.~P., A.~N., A.~M., and E.~D. were supported by the 
U.S. Department of Energy (DOE),
Office of Science, Basic Energy Sciences (BES), Materials Science and
Engineering Division. G.~A. and S.~J. were supported by 
the Scientific Discovery through Advanced Computing (SciDAC) program funded by
the U.S. Department of Energy, Office of Science, Advanced Scientific Computing
Research and Basic Energy Sciences, Division of Materials Sciences and
Engineering. 
N.~D.~P. was also partially supported by the National
Science Foundation Grant No. DMR-1404375.
Part of this work was conducted at the
Center for Nanophase Materials Sciences, sponsored by the
Scientific User Facilities Division (SUFD), BES, DOE, under
contract with UT-Battelle. 
Computer time provided in part by resources supported
by the University of Tennessee and Oak Ridge National
Laboratory Joint Institute for Computational Sciences.

{\bf  Author contributions} --- 
N. D. Patel and E. Dagotto planned the project. N. D. Patel performed all the DMRG calculations for multi-orbital Hubbard model. A. Nocera and G. Alvarez developed the DMRG++ computer program. E. Dagotto, A. Moreo and S. Johnston provided important insight into the understanding of orbital excitations, and comparison with RIXS experiment.

{\bf  Competing interests} --- 
The authors declare no competing interests.

\begin{widetext}
\clearpage
\begin{center}
\textbf{\large Supplemental: Fingerprints of an Exotic Orbital-Selective Mott Phase \\ 
in the Block Magnetic State of BaFe$_2$Se$_3$ Ladders}
\end{center}
\end{widetext}

\setcounter{equation}{0}
\setcounter{figure}{0}
\setcounter{table}{0}
\setcounter{page}{1}
\makeatletter
\renewcommand{\theequation}{S\arabic{equation}}
\renewcommand{\thefigure}{S\arabic{figure}}
\renewcommand{\bibnumfmt}[1]{[S#1]}

\section{Operators and Observables}

In this section, we define all the operators that are used in the main text:

%
\begin{equation} 
\begin{split}
n_{i\gamma\sigma} &= c^{\dagger}_{i \gamma \sigma} c_{i \gamma \sigma},  \\
n_{i\gamma} &= n_{i \gamma \up} + n_{i \gamma \dn}, \\
S^{\kappa}_{i \gamma} &=\frac{1}{2} \sum_{\sigma,\sigma'} c^{\dagger}_{i \gamma \sigma} \sigma^{\kappa}_{\sigma \sigma'} c_{i \gamma \sigma'},  \\
P_{i \gamma} &= c_{i \gamma \up} c_{i \gamma \dn},
\end{split}
\end{equation}
%
where $\gamma$ is the orbital index, $\sigma$ and $\sigma'$ are the spin index, and 
$\sigma^{\kappa}$ are the Pauli matrices with $\kappa=\{x,y,z\}$ being Cartesian components.
The average occupation and charge fluctuations (Fig.1 of main text) are
defined as
%
\begin{equation} 
\begin{split}
\< n_{\gamma} \> &= \frac{1}{L} \sum_{i, \sigma} n_{i\gamma\sigma}, \\
\< \delta N_{\gamma}^2 \> &= \frac{1}{L} \sum_{i}  \< n_{i \gamma} \ n_{i \gamma} \> - \< n_{i \gamma} \> \< n_{i \gamma} \>,
\end{split}
\end{equation}
%
where $L$ is the number of sites in the lattice.
The static spin-spin correlations (inset of Fig. 1) is calculated using
\begin{equation} \label{eq:Sk}
S(k) = \frac{1}{L^2} \sum_{i,j} e^{-i k (i - j)} \langle {{\mathbf{S}_i}\cdot{\mathbf{S}_{j}}} \rangle,
\end{equation}
where ${\mathbf{S}_i} = \sum_{\gamma} \mathbf{S}_{i \gamma}$.
In this supplemental, we also show results for $n^{\gamma}_k$ defined as 
\begin{equation} \label{eq:Sk}
\begin{split}
c_{j \gamma} &= c_{j \gamma \up} + c_{j \gamma \dn}, \\
n^{\gamma}(k) &= n^{\gamma}_k = \frac{1}{L^2} \sum_{i,j} e^{-i k (i - j)} \langle c^{\dagger}_{i \gamma} c_{j \gamma} \rangle.
\end{split}
\end{equation}

\subsection{Spectral functions and sum rules}
To characterize the OSMP, the dynamical response functions (shown below) are calculated 
using DMRG 
%
\begin{equation} \label{eq:Greenijw}
\begin{split}
O(i,j,\omega) &= \frac{-1}{\pi} \text{Im} \big[ \langle \psi_0 | O^{\dagger}_{i} \frac{1}{\omega - H + E_g + {\mathrm i}\eta} O_{j} | \psi_0 \rangle \big],
\end{split} 
\end{equation}
%
where the local operator $O_i$ can represent any degree of freedom of the model. 
In general, these 
functions are Fourier transformed into the crystal momentum domain to calculate the 
momentum-energy resolved spectra that is relevant to experiments: 
%
\begin{equation} \label{eq:Greenkw}
\begin{split}
O(k,\omega) &=  \frac{1}{L^2} \sum_{i,j} e^{-ik(i-j)} O(i,j,\omega). \\
\end{split}
\end{equation}
%
Note that within DMRG, the site $j$ is fixed to the center of the 
lattice ($d=L/2-1$) to reduce the edge effects and computational cost, and therefore the modified 
Fourier transform becomes  
%
\begin{equation} 
\begin{split}
O(k,\omega) &=  \frac{1}{L} \sum_{i} e^{-ik(i-d)} O(i,d,\omega).
\end{split}
\end{equation}
%
Additionally, we use open boundary conditions in the DMRG simulation and therefore 
the quasi crystal-momenta are defined as 
%
\begin{equation} 
\begin{split}
k = \frac{\pi n}{L+1} \ \ \ \text{where} \ \ \ n = 1,2 ... \ L.
\end{split}
\end{equation}
%

\section{Orbital Operators}

The dominant orbitals of an iron atom in iron-based superconductors 
are the five $3d$ orbitals, 
corresponding to well-known $L=2$ orbital angular momenta. 
The corresponding operators $\{L_x,L_y,L_z\}$ are written in the basis of $L_z = \{-2,-1,0,1,2\}$, forming 
$5\times5$ matrices:
%
\[
L_x = \frac{1}{2}
\begin{bmatrix}
    0 & 2 & 0 & 0  & 0 \\
    2 & 0 & \sqrt{6} & 0 & 0 \\
    0 & \sqrt{6} & 0 & \sqrt{6} & 0 \\
    0 & 0 & \sqrt{6} & 0  & 2 \\
    0 & 0 & 0 & 2  & 0 \\
\end{bmatrix}
\]
\[
L_y = \frac{-{\mathrm i}}{2}
\begin{bmatrix}
    0 & 2 & 0 & 0  & 0 \\
    -2 & 0 & \sqrt{6} & 0 & 0 \\
    0 & -\sqrt{6} & 0 & \sqrt{6} & 0 \\
    0 & 0 & -\sqrt{6} & 0  & 2 \\
    0 & 0 & 0 & -2  & 0 \\
\end{bmatrix}
\]
\[
L_z = 
\begin{bmatrix}
    2 & 0 & 0 & 0 & 0 \\
    0 & 1 & 0 & 0 & 0 \\
    0 & 0 & 0 & 0 & 0 \\
    0 & 0 & 0 & -1 & 0 \\
    0 & 0 & 0 & 0 & -2 \\
\end{bmatrix}
\]
%
DMRG is a real-space algorithm, and therefore we must write 
these matrices in the orbital basis using the transformation
%
\begin{equation}
\begin{split}
|-2\> &= \frac{1}{\sqrt{2}} \Big( |x^2 - y^2\> - {\mathrm i} |xy\> \Big), \\
|-1\> &= \frac{1}{\sqrt{2}} \Big( |xz\> - {\mathrm i} |yz\> \Big), \\
|0\> &= |z^2\>, \\
|1\> &= \frac{-1}{\sqrt{2}} \Big( |xz\> + {\mathrm i} |yz\> \Big), \\
|2\> &= \frac{1}{\sqrt{2}} \Big( |x^2 - y^2\> + {\mathrm i} |xy\> \Big), \\
\end{split}
\end{equation}
%
where $\{ |x^2 - y^2\>, |z^2\>, |xz\>, |yz\>, |xy\> \}$ are the 
five iron $3d$ orbitals. The operators $\{L_x,L_y,L_z\}$ represented in 
the orbital basis are:
%
\[
L_x = \frac{{\mathrm i}}{2}
\begin{bmatrix}
    0 & 0 & 0 & 1  & 0 \\
    0 & 0 & 0 & \sqrt{3} & 0 \\
    0 & 0 & 0 & 0 & 1 \\
    -1 & -\sqrt{3} & 0 & 0  & 0 \\
    0 & 0 & -1 & 0  & 0 \\
\end{bmatrix}
\]

\[
L_y = \frac{{\mathrm i}}{2}
\begin{bmatrix}
    0 & 0 & -1 & 0  & 0 \\
    0 & 0 & \sqrt{3} & 0 & 0 \\
    1 & -\sqrt{3} & 0 & 0 & 0 \\
    0 & 0 & 0 & 0  & 1 \\
    0 & 0 & 0 & -1  & 0 \\
\end{bmatrix}
\]

\[
L_z = {\mathrm i}
\begin{bmatrix}
    0 & 0 & 0 & 0 & 2 \\
    0 & 0 & 0 & 0 & 0 \\
    0 & 0 & 0 & 1 & 0 \\
    0 & 0 & -1 & 0 & 0 \\
    -2 & 0 & 0 & 0 & 0 \\
\end{bmatrix}.
\]
%
In the main text, we employ the iron $t_{2g}$ orbitals, 
{\it i.e.,} we drop the contribution from $|x^2 - y^2\>, |z^2\>$ 
orbitals. After this approximation, we obtain the local orbital angular momentum 
operators used in the main text (site index understood)
%
\begin{equation}
\begin{split}
L^{x} &= {\mathrm i} (\cd{xz} \c{xy}^\pdag - \cd{xy} \c{xz}^\pdag), \\
L^{y} &= {\mathrm i} (\cd{yz} \c{xy}^\pdag - \cd{xy} \c{yz}^\pdag), \\
L^{z} &= {\mathrm i} (\cd{xz} \c{yz}^\pdag - \cd{yz} \c{xz}^\pdag).
\end{split}
\end{equation}
%

\section{Additional Results}

In this section, we show tests performed to ensure the quality of the
presented results. We first show $n^{\gamma}(k)$ that is in 
agreement with previous studies with a similar model~\cite{SLiOSMP}.
A significant change in $n^{a/b}(k)$ indicates metallic behavior 
of these orbitals. On the contrary, $n^{c}(k)$ shows little variation in 
$k$ that is associated with a gapped orbital $c$. 

%
\begin{figure}[thbp]
\begin{center}
\hspace{-0.0cm}
 \begin{overpic}[trim = 0mm 0mm 0mm 0mm,
height=0.28\textwidth,width=0.38\textwidth,angle=0]{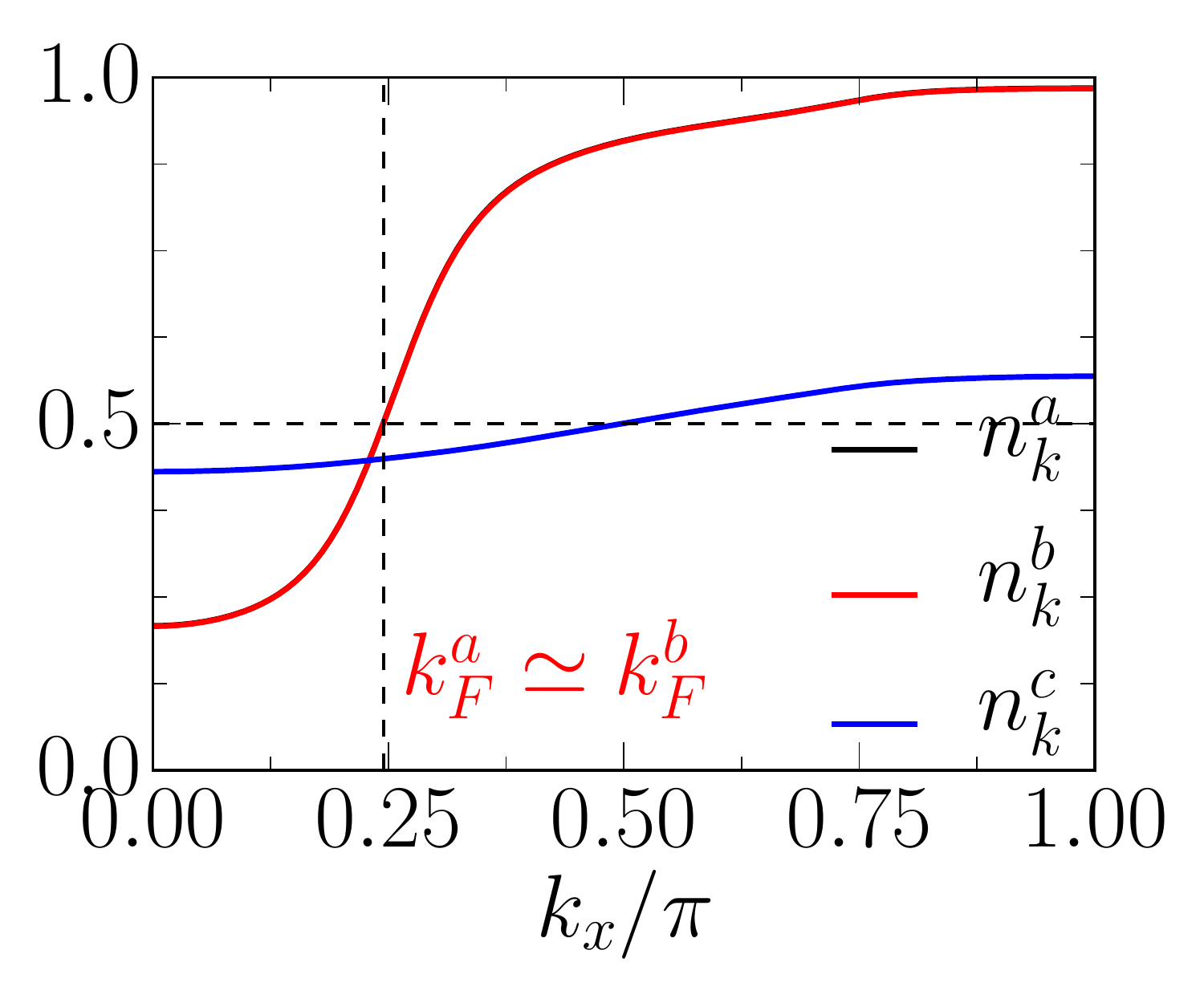}
\end{overpic}
\end{center}
\vspace{-0.5cm}
\caption{(color online) 
$n^\gamma(k)$ for the Block OSMP ($U/W=0.8$) using a $32$ sites chain.
The largest increase in $n^\gamma(k)$ of the itinerant orbitals $a$ (black) and $b$ (red) 
occurs at $k/\pi \simeq 0.25$, i.e. the Fermi momentum. Note that results for $a$ and $b$ are almost
identical, thus indistinguishable in the figure. $n^c(k)$ of the 
localized orbital $c$ has only a slight variation in $k$, 
suggesting a gap in the single-particle states of orbital $c$. 
These results are in agreement with previous studies 
using Quantum Monte Carlo~\cite{SLiOSMP}.
}
\label{fig:S1}
\end{figure}
%

Furthermore, all spectral quantities defined by 
equations~\ref{eq:Greenijw} and \ref{eq:Greenkw} must satisfy a sum rule. 
In general, it can be shown that integrating the spectral function over momentum $k$ and energy $\w$ 
gives a quantity related to a unique static observable. As an example, 
we use the single-particle spectral function $A^c(k,\w)$ of orbital $c$ below (above) the 
Fermi energy ($E_F$) representing the filled (unfilled) electron (hole) states. 
The electron and hole components of $A^c(k,\w)$ are defined as 
%
\begin{equation} \label{eq:Akw}
\begin{split}
A_e^c(k,\omega) &= \frac{-1}{\pi L^2} \times  \\ \sum_{i,j} \text{Im} \big[ &\langle \psi_0 | c^{\dagger}_{ic} \frac{1}{\omega - H + E_g + {\mathrm i}\eta} c_{jc} | \psi_0 \rangle \big] e^{-ik(i-j)}, \\
\\
A_h^c(k,\omega) &= \frac{-1}{\pi L^2} \times \\ \sum_{i,j} \text{Im} \big[ &\langle \psi_0 | c_{ic} \frac{1}{\omega + H + E_g + {\mathrm i}\eta} c^{\dagger}_{jc} | \psi_0 \rangle \big]  e^{-ik(i-j)}, 
\end{split}
\end{equation}
%
where $c$ is the orbital index and $c_{jc} = c_{jc\up} + c_{jc\dn}$. 
To obtain the sum rule of this quantity, we first sum over the momentum $k$. 
This sum simply results in $L \delta_{i,j}$, giving us the local response that is the
single-particle density of states,
%
\begin{figure}[h]
\begin{center}
 \begin{overpic}[trim = 1.2cm 0mm 0mm 0mm,
height=0.28\textwidth,width=0.32\textwidth,angle=0]{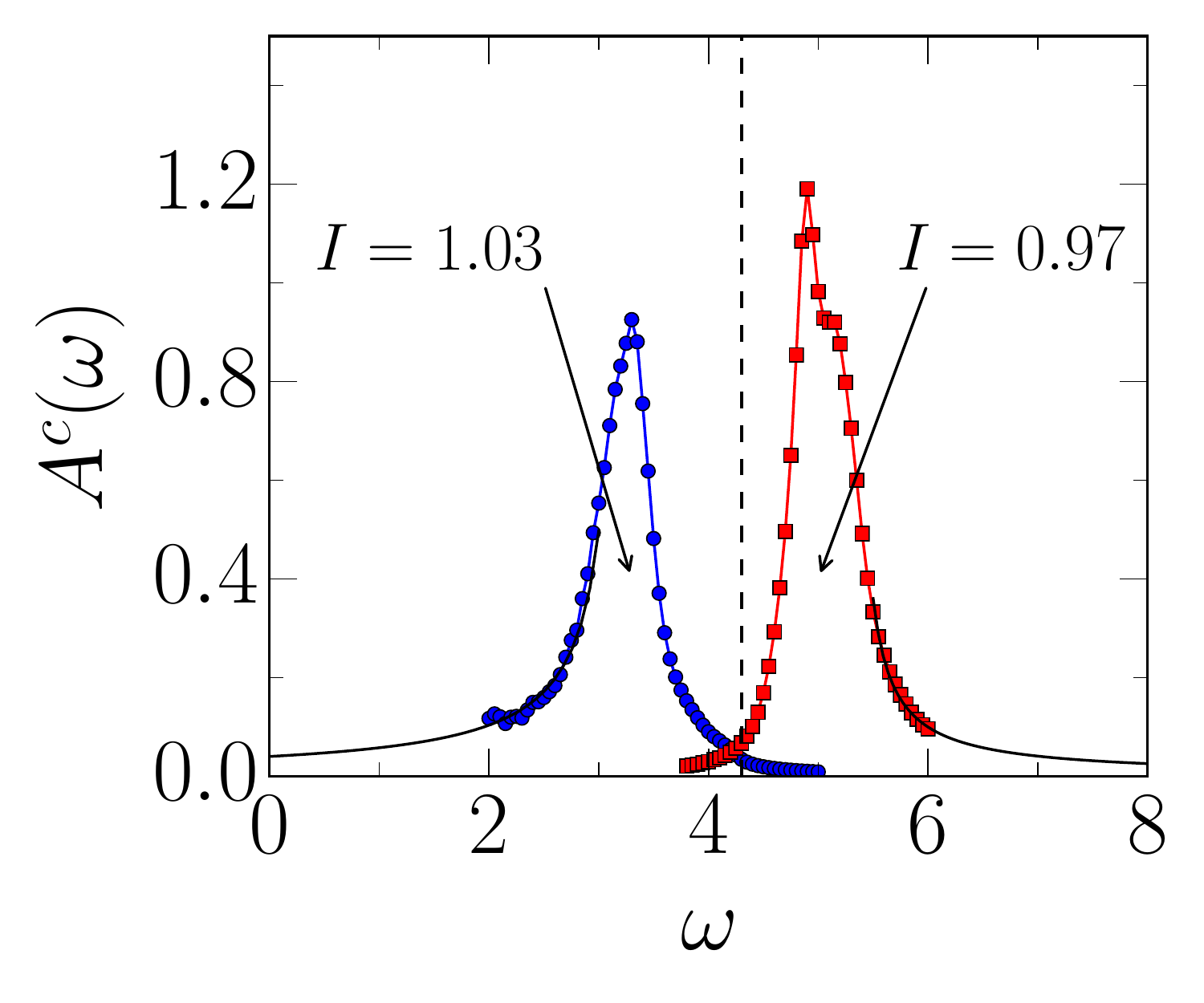}
\end{overpic}\hspace{-0.25cm}
\end{center}
\vspace{-0.5cm}
\caption{(color online) Density of states of the Mott orbital $c$, calculated using 
the electron ($A^c_e$, blue) and hole ($A^c_h$, red) components, employing $24$ sites, 
$\Delta \w = 0.05,$ and $\eta=0.1$. The vertical grey line represents the 
Fermi energy $E_F \simeq 4.3$~eV. The 
tails of the density of states 
are fitted in order to account for the missing tail weights. 
$I \simeq 1$ is the integrated value of the electron and hole part of 
the density of states which is respectively equal to the average local electron and hole 
occupations. 
}
\label{fig:S2}
\end{figure}
%
%
\begin{equation} \label{eq:Aw}
\begin{split}
A_e^c(\omega) &=  -\frac{1}{\pi L} \sum_{i} \text{Im} \big[ \langle \psi_0 | c^{\dagger}_{ic} \frac{1}{\omega - H + E_g + {\mathrm i}\eta} c_{ic} | \psi_0 \rangle \big], \\
A_h^c(\omega) &=  -\frac{1}{\pi L} \sum_{i} \text{Im} \big[ \langle \psi_0 | c_{ic} \frac{1}{\omega + H + E_g + {\mathrm i}\eta} c^{\dagger}_{ic} | \psi_0 \rangle \big],
\end{split}
\end{equation}
%
shown in Figure~\ref{fig:S2}. Further integration over $\w$ of the imaginary part 
(Lorentzian poles) of the electron (hole) part simply gives the total electron (hole) 
density of orbital $c$:
%
\begin{equation} \label{eq:Aw}
\begin{split}
n_e^c &=  \frac{1}{L} \sum_{i} \langle \psi_0 | c^{\dagger}_{i} c_{i} | \psi_0 \rangle , \\
n_h^c &=  \frac{1}{L} \sum_{i} \langle \psi_0 | c_{i} c^{\dagger}_{i} | \psi_0 \rangle .
\end{split}
\end{equation}
%
This is explicitly done within our calculations by integrating over the density of 
states (Fig.~\ref{fig:S2}). The integration of the electron (blue curve) 
and hole (red curve) portions lead to approximately $1.0$ that is consistent, within the accuracy
of our results, with calculations 
of the local density from the ground-state. Note that a singly occupied orbital is 
one of the characteristics of a Mott insulator.

\begin{figure}[t]
\vspace{0.1cm}
\begin{center}
 \begin{overpic}[trim = -0cm 0.8cm 1.2cm 0.0cm,
height=0.28\textwidth,width=0.32\textwidth,angle=0]{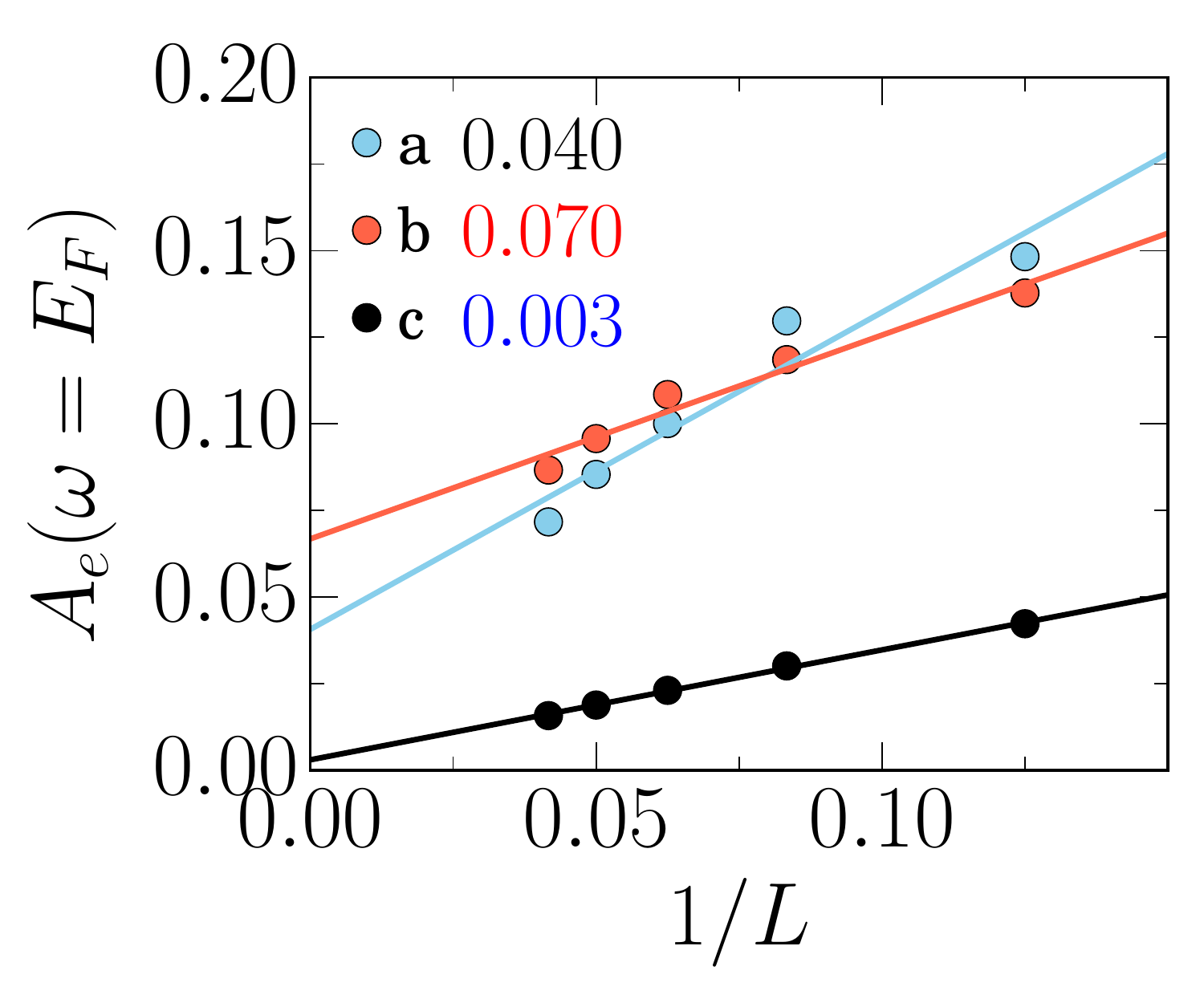}
\end{overpic}
\end{center}
\vspace{-0.14cm}
\caption{(color online) Finite-size scaling of the orbital-resolved electron part of the 
density of states at the Fermi energy ($E_F$). The quasi-particle 
weight of the Mott orbital approaches $0$ (more accurately, $0.003$) with increasing 
system size while the weight remains finite for the itinerant 
orbitals ($0.04$ and $0.07$). Note that $A_e(E_F)$ of the itinerant 
orbitals $a$ and $b$ is an order magnitude larger than the Mott orbital $c$, further 
emphasizing that orbital $c$ is a Mott insulator.
}
\label{fig:S3}
\end{figure}

We also performed finite-size scaling on the density of states at the Fermi energy $E_F \simeq 4.3$~eV. 
Figure \ref{fig:S3} shows that the $L \rightarrow \infty$ extrapolated quasi-particle weight, 
$A(\w=E_F)$, of orbital $c$ is an order of magnitude smaller than the
almost degenerate orbitals $a$ and $b$. The near zero weight of orbital $c$ at the Fermi 
energy is consistent with a Mott phase, providing further evidence of the presence of an OSMP 
as the ground-state.
%
%

\section{Reproducing data using DMRG++}
The full open source code, sample inputs, and corresponding
computational details can be found at
\url{https://g1257.github.io/papers/86/}.

%
%


\end{document}